\documentclass[aps,amsmath,amssymb,nofootinbib,superscriptaddress,showpacs,floatfix,prb,twocolumn]{revtex4-1}

\usepackage{latexsym}
\usepackage{graphicx}
\usepackage{times,psfrag,subfigure}
\usepackage{enumerate}
\usepackage{amsmath}
\usepackage{dsfont}
\usepackage{dcolumn}
\usepackage{bm,bbm}
\usepackage{color}
\usepackage{latexsym,amsmath,amssymb,bm,euscript}
\usepackage{dsfont}
\usepackage{textcomp}
\usepackage{tabularx}
\usepackage{setspace}
\usepackage{ctable}
\usepackage{sidecap}
\usepackage{placeins}
\usepackage{threeparttable}
\usepackage{multirow}
\usepackage[utf8]{inputenc}
\usepackage[english, ngerman]{babel}

\def\T{\mathcal{T}}

\def\half{{1\over 2}}

\def\H{\mathcal{H}}

\def\b0{{\bm0}}
\def\bk{{\bm k}}

\newcommand{\ket}[1]{| #1 \rangle}

\newcommand{\beq}{\begin{equation}}
\newcommand{\eeq}{\end{equation}}
\newcommand{\beqarray}{\begin{eqnarray}}
\newcommand{\eeqarray}{\end{eqnarray}}

\graphicspath{{./Images/}}

\begin{document}
\selectlanguage{english}
\allowdisplaybreaks

\title{Surface floating 2D bands in layered nonsymmorphic semimetals:\\ $\rm \mathbf{ZrSiS}$ and related compounds}

\date{\today}

\author{Andreas Topp}
\email{a.topp@fkf.mpg.de}
\affiliation{Max-Planck-Institut f\"ur Festk\"orperforschung, Heisenbergstrasse 1, D-70569 Stuttgart, Germany}

\author{Raquel Queiroz}
\email{raquel.queiroz@weizmann.ac.il}
\affiliation{Department of Condensed Matter Physics, Weizmann Institute of Science, Rehovot 76100, Israel}
\affiliation{Max-Planck-Institut f\"ur Festk\"orperforschung, Heisenbergstrasse 1, D-70569 Stuttgart, Germany}

\author{Andreas Grüneis}
\affiliation{Max-Planck-Institut f\"ur Festk\"orperforschung, Heisenbergstrasse 1, D-70569 Stuttgart, Germany}
\affiliation{Institute for Theoretical Physics Vienna University of Technology, A-1040 Vienna, Austria}
\author{Lukas Müchler}
\affiliation{Department of Chemistry, Princeton University, Princeton, New Jersey 08544, USA}
\author{Andreas Rost}
\affiliation{Max-Planck-Institut f\"ur Festk\"orperforschung, Heisenbergstrasse 1, D-70569 Stuttgart, Germany}
\author{Andrei Varykhalov}
\affiliation{Helmholtz-Zentrum Berlin f\"ur Materialien und Energie, Elektronenspeicherring BESSY II, Albert-Einstein-Stra\ss e 15, 12489 Berlin, Germany}
\author{Dmitry Marchenko}
\affiliation{Helmholtz-Zentrum Berlin f\"ur Materialien und Energie, Elektronenspeicherring BESSY II, Albert-Einstein-Stra\ss e 15, 12489 Berlin, Germany}
\author{Maxim Krivenkov}
\affiliation{Helmholtz-Zentrum Berlin f\"ur Materialien und Energie, Elektronenspeicherring BESSY II, Albert-Einstein-Stra\ss e 15, 12489 Berlin, Germany}
\author{Fanny Rodolakis}
\affiliation{Argonne National Laboratory, 9700 S. Cass Ave, Argonne, Il 60439}
\author{Jessica McChesney}
\affiliation{Argonne National Laboratory, 9700 S. Cass Ave, Argonne, Il 60439}
\author{Bettina V. Lotsch}
\affiliation{Max-Planck-Institut f\"ur Festk\"orperforschung, Heisenbergstrasse 1, D-70569 Stuttgart, Germany}
\author{Leslie M. Schoop}
\affiliation{Max-Planck-Institut f\"ur Festk\"orperforschung, Heisenbergstrasse 1, D-70569 Stuttgart, Germany}
\author{Christian R. Ast}
\affiliation{Max-Planck-Institut f\"ur Festk\"orperforschung, Heisenbergstrasse 1, D-70569 Stuttgart, Germany}

\begin{abstract}
In this work, we present a model of the surface states of nonsymmorphic semimetals. These are derived from surface mass terms that lift the high degeneracy imposed in the band structure by the nonsymmorphic bulk symmetries. Reflecting the reduced symmetry at the surface, the bulk bands are strongly modified. This leads to the creation of two-dimensional floating bands, which are distinct from Shockley states, quantum well states or topologically protected surface states. We focus on the layered semimetal $\rm ZrSiS$ to clarify the origin of its surface states. We demonstrate an excellent agreement between DFT calculations and ARPES measurements and present an effective four-band model in which similar surface bands appear. Finally, we emphasize the role of the surface chemical potential by comparing the surface density of states in samples with and without potassium coating. Our findings can be extended to related compounds and generalized to other crystals with nonsymmorphic symmetries.
\end{abstract} 

\date{\today}

\pacs{79.60.-i,71.15.Mb,71.20.-b,73.20.-r}

\maketitle

\section{Introduction} 
Surface states have gathered significant interest as they 
reflect both surface details and bulk properties. In view of the current search for nontrivial topology in band insulators \cite{Kane2005quantum,fu2007topological,schnyderPRB08,zhang2009topological,ando2013topological,Nobel2016} and semimetals \cite{wan2011topological,yan2012topological,Fang2012,Chiu2014a,Liu2014DiscoveryNa3Bi.,Liu2014,Shekhar2015a,Chan2016}, it has become crucial to characterize different types of surface states which may emerge in these systems. Specifically, it is 
important to distinguish surface states which reflect surface chemistry, such as dangling bond states \cite{northrup1986origin} or quantum well states \cite{paggel1999quantum}, from surface states that arise from bulk band topology. 
Shockley and Tamm 
states are 
the most common examples of surface states 
\cite{Davison1992}, arising as an additional solution of the Hamiltonian at the interface within a projected band gap. The additional requirement of a band inversion 
implies a nontrivial
band topology 
even 
in the absence of spin-orbit coupling (SOC). 
A strong topological index is expected when a full gap is opened by SOC \cite{yan2015topological}. 
Band inversions justify the persistence of surface states in noble metals even in the presence of absorbents such as alkali metals \cite{lindgren1979energy,tang1993unoccupied,kevan1986anomalous,sandl1994surface}. 
However, these states are generally not robust against back-scattering, since there is no symmetry which protects the energy separation between surface and bulk states.
This is in clear contrast with topological surface states of band insulators such as $\rm Bi_2Se_3$ \cite{zhang2009topological}. 
On the other hand, topologically trivial
surface states can 
arise in quantum wells (thin films) \cite{paggel1999quantum}, due to band bending in semiconductors \cite{benia2011reactive}, or simply from dangling bonds from unbound electrons in semiconductors or insulators \cite{northrup1986origin}. Unlike nontrivial states, trivial states are easily affected, even eliminated, by surface modifications \cite{northrup1986origin,cartier1993passivation}. 
Finally, submonoloayer modifications, such as surface alloying, can lead to a new and (mostly) independent two-dimensional electronic structure resulting in states at the surface as well \cite{ast_giant_2007}.

Recently, much attention has been drawn to a number of nonsymmorphic materials exhibiting extended linear band dispersions and complex Dirac line nodes, along with extremely large magnetoresistance  \cite{Schoop2016,lv2016extremely,ali2016butterfly,topp2016non,Takane2016a,xu2015two}. The most prominent representative is the ternary compound ZrSiS, which has been shown to exhibit an unusual surface electronic structure, which coexists with bulk bands near the Fermi level \cite{Schoop2016}.
Despite various proposals based on $\rm ZrSiS$ and $\rm ZrSnTe$ \cite{Neupane2016ObservationZrSiS,lou2016emergence}, 
a quantitative account of the origin and topological character of recurring surface states in this class of compounds still remains unresolved. 

\begin{figure}[b]
    \centering
    \includegraphics[width=0.5\textwidth]{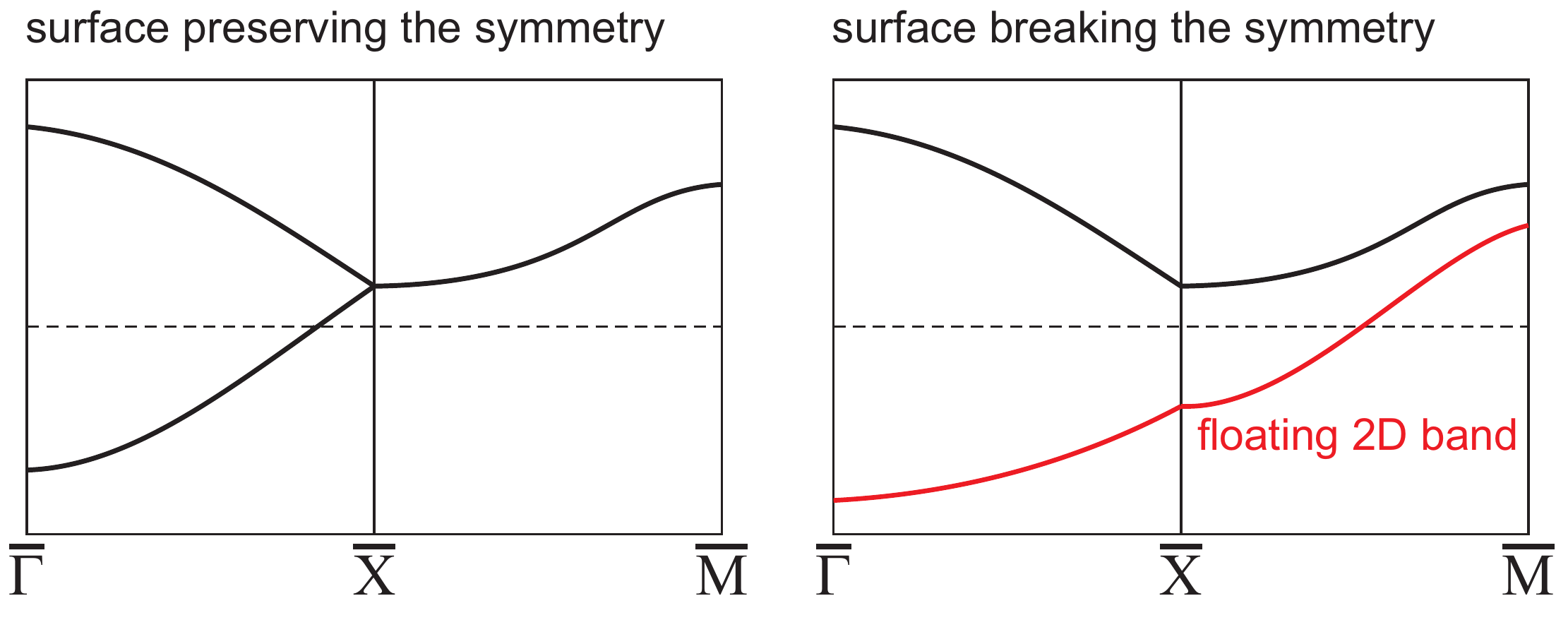}
    \caption{
    Schematic of the surface states possible in nonsymmorphic materials, as occurrent in $\rm ZrSiS$.
    }
    \label{fig:Schematic}
\end{figure}

In this work, we 
propose that the surface states of $\rm ZrSiS$ are not derived from the nonsymmorphic bulk band topology, such as hourglass\cite{wang2016hourglass} and Dirac \cite{wieder2017wallpaper} surface states, nor are they due to a structural surface modification, but 
instead are the result of a reduced 
symmetry at the surface.

As for most nonsymmorphic space groups, the bulk electronic structure of the tetragonal space group $P4/nmm$ (No.\ 129) is characterized by high band degeneracies (up to four-fold in this case) at certain high symmetry momenta.  
At the surface, translation symmetry along one direction is broken, and consequently the symmetry is reduced. 
In the present layered structure, the natural cleavage plane (001) has its symmetry reduced to the symmorphic wallpaper group $P4mm$ (No. 99). Consequently, the high band degeneracies are no longer protected and can be lifted, which can be seen schematically in Fig.\,\ref{fig:Schematic}.

In this regard, the surface state of $\rm ZrSiS$ and related compounds can be seen as `floating' two-dimensional bands, localized at the surface. Such a symmetry derived surface electronic structure does not fit any of the aforementioned surface state models, 
and is not particular to this space group. Surface floating bands are 
a common occurrence in nonsymmorphic semimetals, which feel a reduced symmetry at the surface.
Since they originate from bulk bands they can not be saturated by adsorbents or show a reduced cross section for high photon energies. 

This article is organized as follows. We 
present the methods in Sec.\,\ref{sec:Methods}, 
In Sec.\,\ref{sec:Overview} 
we present angle-resolved photoemission spectroscopy (ARPES) data at various photon energies, 
and slab density functional theory (DFT) calculations. 
In Sec.\,\ref{sec:theo} we show a minimal tight-binding model that qualitatively 
describes the origin of the 
floating two-dimensional surface bands. Finally, in  
Sec.\,\ref{sec:add}, we study the sensitivity of the surface bands to 
surface details 
by presenting ARPES and DFT calculations of $\rm ZrSiS$ with 
a monolayer of potassium evaporated on the surface.

\section{Methods}\label{sec:Methods}

For the ARPES measurements,  crystals were cleaved in-situ and measured in ultra-high vacuum (low $10^{-10}$\,mbar range). Low energy ARPES spectra were recorded with the $1^2$-ARPES experiment installed at the UE112-PGM2a beamline at BESSY-II. The measurement temperature was 40\,K.
Soft X-ray ARPES measurements (between $h\nu=248\,$eV and 1000\,eV) were performed at room temperature and 77\,K at the 29ID-IEX beamline (Advanced Photon Source, Argonne National Laboratory) using a hemispherical Scienta R4000 electron analyzer with a pass energy of $200$\,eV (energy and angular resolution are 220\,meV and 0.1$^\circ$, respectively). 

For the periodic DFT calculations we employed the Vienna $ab-initio$ simulation package (\texttt{VASP}) \cite{kresse1999} in the framework of the projector augmented wave (PAW) method.
For Zr the $4s$, $4p$, $5s$ and $4d$ were treated as valence states, whereas for S and Si the $3s$, $3p$ states were treated as valence states. For K coating, we employed a PAW potential that treats the $3s$, $3p$ and $4s$ as valence states.
The kinetic energy cutoff for the plane wave basis was set to 500\,eV. Unless stated otherwise, we have employed the Perdew-Burke-Ernzerhof (PBE) exchange correlation energy functional. The Brillouin zone integrations were carried out using  a 4~$\times$~4~$\times$~1 $\Gamma$-centered $k$-mesh.
The surface model has a slab thickness of seven unit cells, while the original publication on $\rm ZrSiS$ used a thickness of five unit cells \cite{Schoop2016}. The calculations show no essential difference in comparison. The atomic positions have been kept fixed to the experimental bulk positions. For the adsorption of the K atoms, we have kept the atomic positions of the atoms in the surface fixed, allowing only the K atoms to relax. A coverage of one K atom per surface unit was used and applied to both sides of the slab.

\section{Surface state: ARPES and DFT}\label{sec:Overview}

In this section, we present ARPES measurements of the surface state of {\rm ZrSiS} and show its appearance in a slab DFT calculation. We show a remarkably good correspondence between the two, and present a study of the orbital character of the surface state. 

\begin{figure}
    \centering
    \includegraphics[width=0.5\textwidth]{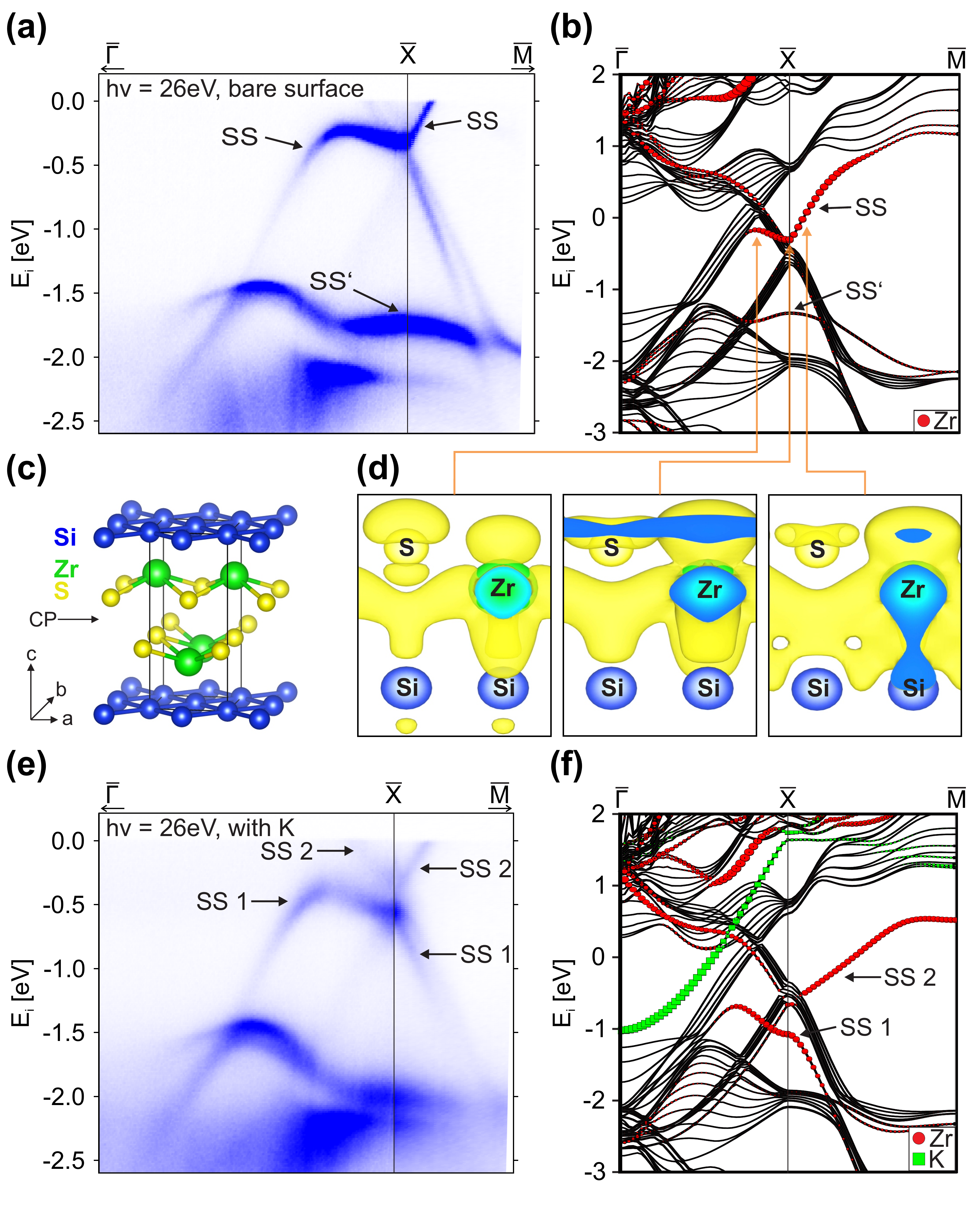}
    \caption{Visualization of the surface states in $\rm ZrSiS$. (a) and (b) show dispersion plots along $\mathrm{\overline{\Gamma}\,\overline{X}\,\overline{M}}$. The path through the Fermi surface is shown in red in Fig.\,\ref{fig:ARPES_FS}. (a) ARPES data of a bare surface, presenting the surface states (SS and SS') as high intensity features crossing the bulk bands. (b) DFT slab calculations where the surface state character is indicated by the size of the red circles. (c) Crystal structure of $\rm ZrSiS$. The unit cell is shown in black. The cleavage plane (CP) is indicated by a black arrow. (d) Isosurfaces of the charge density of the surface state at different $k$-points. The arrows indicate the corresponding independent particle states in the band structure.
    Only the top-most layer of the supercell is shown because the charge density vanishes in the bulk.
    At the $\mathrm{\overline{X}}$ point the surface state is mostly composed of Zr $d_{z^2}$ and S $p$ states. We find that the dispersion of the surface state towards $\overline{\Gamma}$ and $\mathrm{\overline{M}}$ is linked to a change in Zr $d_{z^2}$, $d_{xz}$ and $d_{yz}$ character.
    (e) and (f) show dispersion plots along $\mathrm{\overline{\Gamma}\,\overline{X}\,\overline{M}}$ for a potassium covered surface. Again, the red circles show the states arising from the surface termination in the calculations. The green squares in (f) show states that originate from an ordered array of potassium atoms on both sides of the slab. In (f), the relaxed $z$-distance between K and the surface is 2.74\,\AA. The potassium mixes the orbital character of the surface bands, resulting in a gapping of the surface related bands (SS1 and SS2) in (e), which is reproduced, but overestimated, in the DFT calculations.
    }
    \label{fig:SS_Overview}
\end{figure}

\begin{figure}
    \centering
    \includegraphics[width=0.45\textwidth]{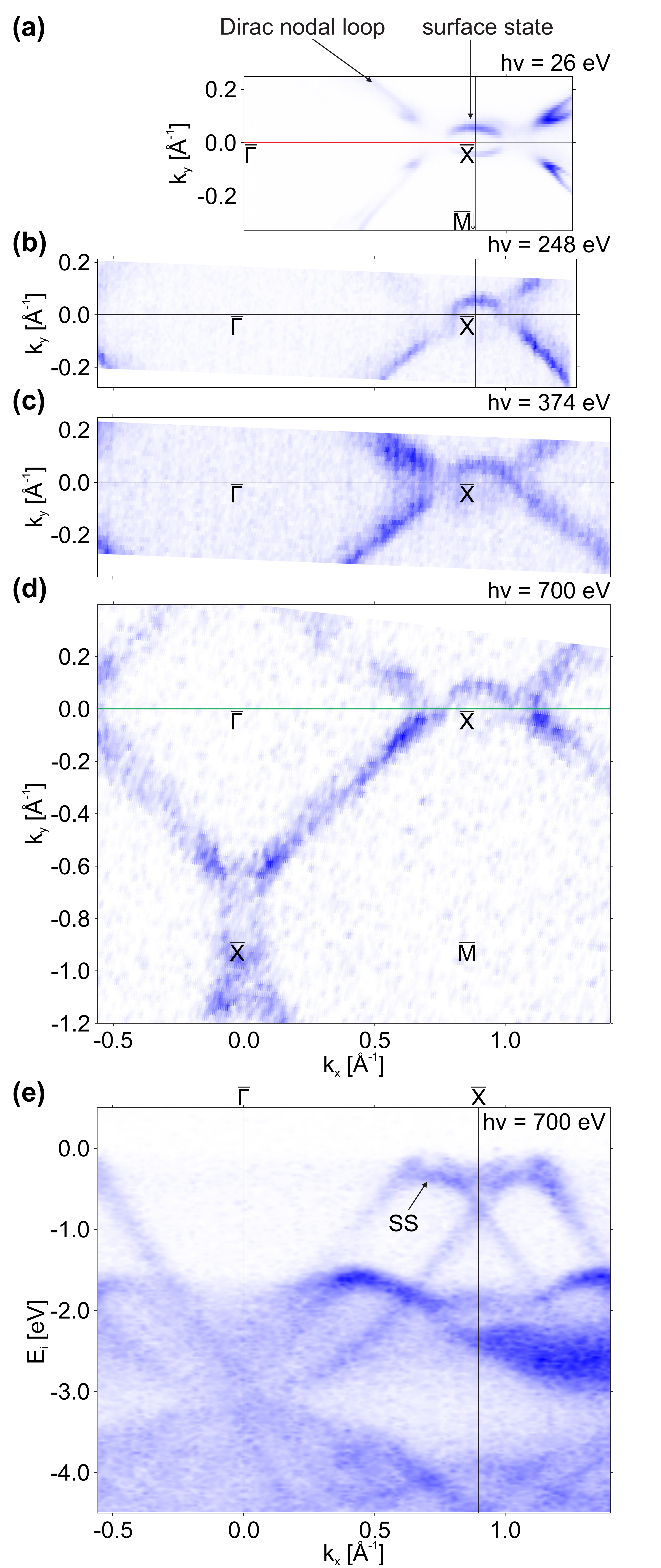}
    \caption{Constant energy plots at the Fermi level for photon energies (a) $h\nu = 26\,$eV, (b) $h\nu = 248\,$eV, (c) $h\nu = 374$\,eV and (d) $h\nu = 700$\,eV. (a) The bare $\rm ZrSiS$ surface shows the right quadrant of the diamond shaped Fermi surface as well as the surface-derived pockets around $\mathrm{\overline{X}}$. The path for the dispersion data in Fig.\,\ref{fig:SS_Overview} is shown in red. (b), (c) and (d) show a similar surface for photon energies of 248\,eV, 374\,eV or 700\,eV respectively. (e) Dispersion along $\mathrm{\overline{\Gamma}\,\overline{X}}$ for $h\nu = 700$\,eV (green line in (d)). The pockets around $\mathrm{\overline{X}}$ are clearly visible in the Fermi surface and the dispersion, showing the considerable weight of these surface states even for high photon energies.}
    \label{fig:ARPES_FS}
\end{figure}

Figure \ref{fig:SS_Overview}(a) shows the experimental band structure of $\rm ZrSiS$ measured at a photon energy of 26\,eV. The dispersion is shown along the $\mathrm{\overline{\Gamma}\,\overline{X}\,\overline{M}}$ line (the path through the surface Brillouin zone (BZ) is shown in red in Fig.\,\ref{fig:ARPES_FS}(a), while the whole BZ can be seen in Fig.\,\ref{fig:effective}(a)). 
At the $\mathrm{\overline{X}}$ point, the linearly dispersing bulk bands are degenerate at $E_\text{i}\approx -0.5$\,eV due to the nonsymmorphic symmetry, and continue to lower initial state energies along the $\mathrm{\overline{X}\,\overline{M}}$ direction. This bulk band is intersected by a very intense surface band (SS) just above the nonsymmorphically degenerate point. A second very intense surface band (SS') is observed at lower initial state energy ($E_\text{i}\approx -1.7$\,eV). In the following, we will focus the discussion on the surface band closer to the Fermi level, although the same arguments hold for the lower surface band. Comparing the experimental results with theoretical calculations in Fig.\,\ref{fig:SS_Overview}(b), we find good agreement with the dispersion of both surface bands. The surface bands with strong Zr character are highlighted by red circles (larger radius indicates stronger Zr character).

To underline the surface character of the states described above, Fig.\,\ref{fig:SS_Overview}(d) shows the calculated charge density isosurfaces of the surface state. The isosurface is shown in real space in the first half unit cell below the cleavage plane (indicated by an arrow in Fig.\,\ref{fig:SS_Overview}(c)). The extended blue area stems from the inner side of the isosurface, where it cuts the pictured cell. The figure shows that the corresponding independent particle state localizes at the surface and exhibits mostly Zr $d_{z^2}$ and S $p$ character at the $\mathrm{\overline{X}}$ point. 
The dispersion towards $\overline{\Gamma}$ and $\mathrm{\overline{M}}$ is linked to a change in its Zr $d_{z^2}$, $d_{xz}$ and $d_{yz}$ character.
The band character is determined using \texttt{VASP} by calculating the absolute value of the overlap between the respective state and spherical harmonics that are nonzero inside a radius around the considered atomic center~\footnote{see VASP manual: www.vasp.at .}. This allows us to measure and plot the surface state character for the present band structure diagrams.
A similar statement can be made for the identification of bands with surface character in the case of potassium coated surfaces in Fig.\,\ref{fig:SS_Overview}(e) and (f), which will be discussed in Sec.\,\ref{sec:add}.

The DFT slab calculations are in an overall agreement with the ARPES data, aside from the energy distance between the nonsymmorphic point and the lower surface band at the $\mathrm{\overline{X}}$ point at $E_i=-1.7$\,eV in Fig.\,\ref{fig:SS_Overview}(a), which is underestimated. The very expensive use of a hybrid functional around the $\mathrm{\overline{X}}$ point, however, shows that this distance is in fact increasing with the computationally more expensive method.

We performed additional measurements near the $\mathrm{\overline{X}}$ point at the APS beamline using higher photon energies. Fig.\,\ref{fig:ARPES_FS}(a) shows a comparison of the Fermi surface, measured at Bessy II with 26\,eV photons. The red lines indicate the high symmetry directions $\mathrm{\overline{\Gamma}\,\overline{X}}$ and $\mathrm{\overline{X}\,\overline{M}}$, which correspond to the band dispersions shown in Fig.\,\ref{fig:SS_Overview}. Fig.\,\ref{fig:ARPES_FS}(b), (c) and (d) show the Fermi surface measured at photon energies of 248\,eV, 374\,eV and 700\,eV, respectively. All four Fermi surfaces show the diamond-like Dirac nodal loop around the $\overline{\Gamma}$ point and the surface state as a ring-like structure around the $\mathrm{\overline{X}}$ point. Note that the surface states are visible for very high photon energies, a cut along $\mathrm{\overline{\Gamma}\,\overline{X}}$ in Fig.\,\ref{fig:ARPES_FS}(e) shows them clearly even for 700\,eV in the dispersion, implying that they have considerable cross section in an energy range where many surface states loose their intensity, such that usually only the bulk states are visible \cite{hsieh_resonances_1987,lobo-checa_effect_2011}. We relate this to the origin of these surface states. Since they arise from bulk bands, they still retain some bulk-like behavior concerning their photon energy dependent properties (see Sec.\,\ref{sec:theo}). Higher photon energies do however, increase the inelastic mean free path for electrons from the bulk leading to a relative decrease of the strongly surface localized surface states in comparison to the bulk states. When we compare Fig.\,\ref{fig:SS_Overview}(a) with Fig.\,\ref{fig:ARPES_FS}(d), the dominating surface state intensity is showing exactly such a behaviour.

From the orbital analysis and a similarity of the surface state and the bulk bands, we conclude that the surface state does not have its origin in a band inversion (because it does not start at a Dirac point). Since it is neither due to surface reconstruction or alloying \cite{Schoop2016}, another reason must be found that explains the relationship between the surface states and the bulk band structure. 


\section{Symmetries and effective model} \label{sec:theo}
The matching DFT slab calculations and ARPES measurements confirm the observation of surface states in $\rm ZrSiS$, but do not reveal 
their origin. 
\begin{figure*}
    \centering
    \includegraphics[scale=0.35]{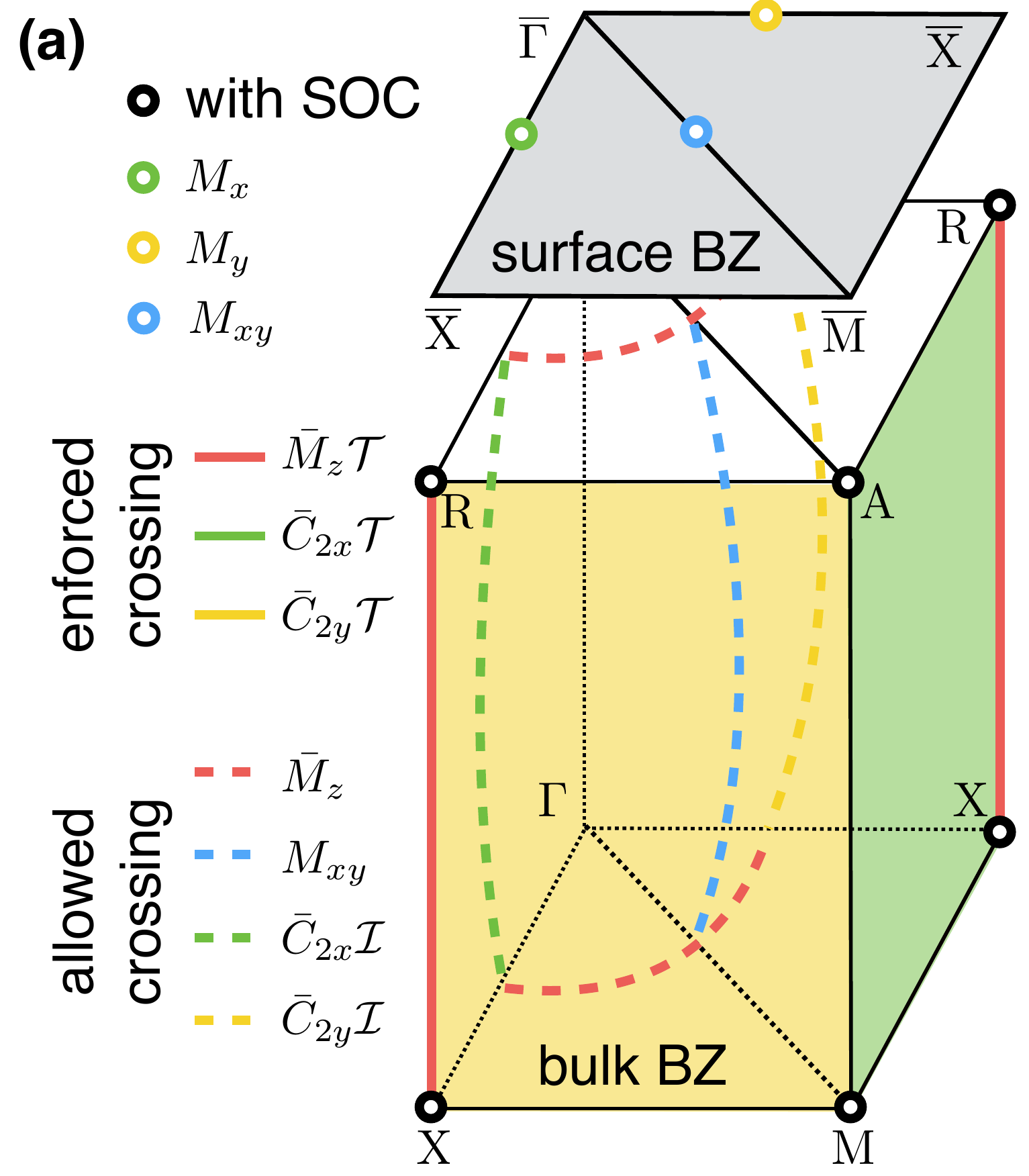}
    \includegraphics[scale=0.35]{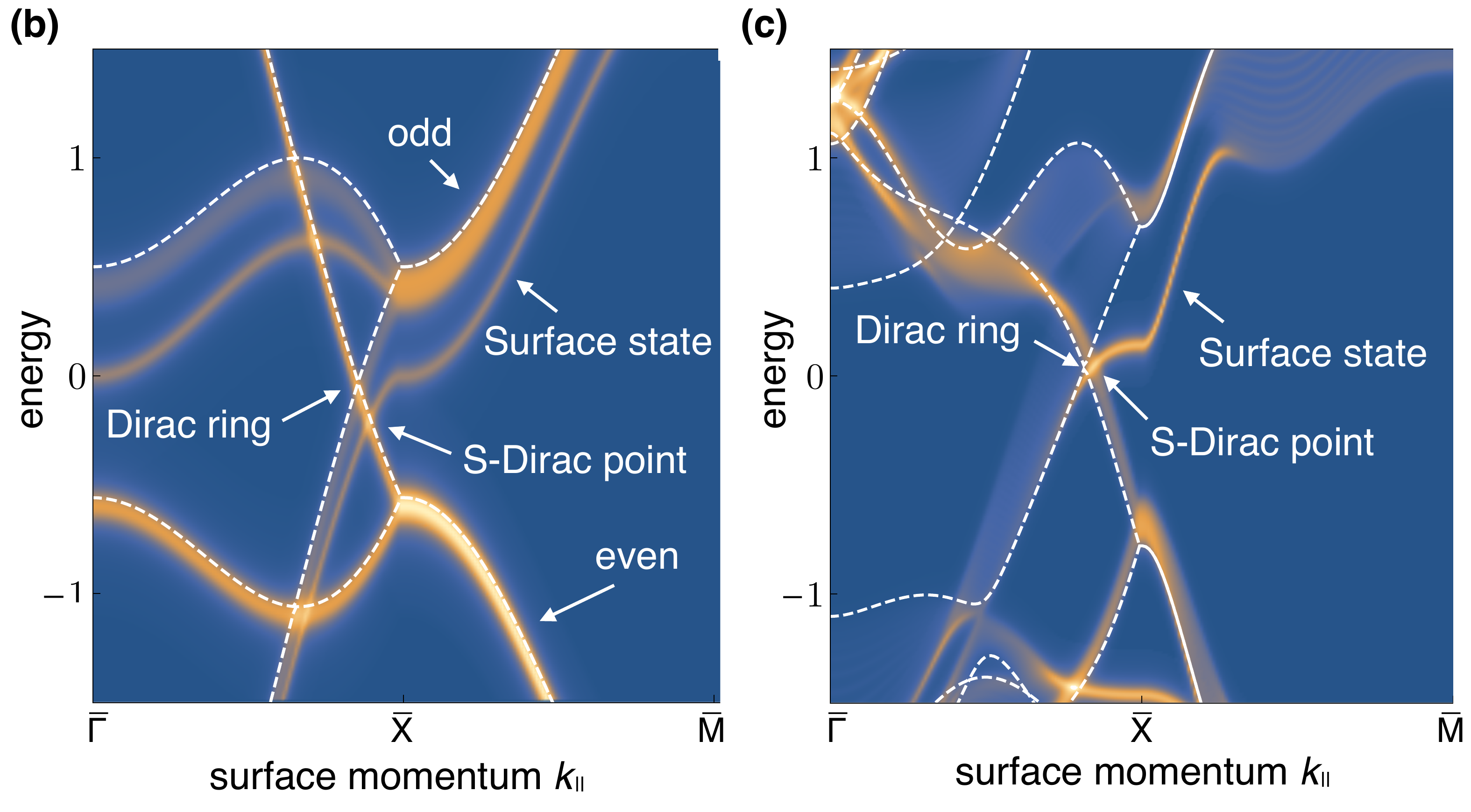}
    \caption{
    (a) Scheme of the symmetries of space group $P4/nmm$. Colored planes, lines, points indicate the symmetry protected degeneracies in the absence of spin-orbit coupling. Solid colors identify momenta where all states are doubly degenerate due to a combination of a nonsymmorphic symmetry and time-reversal; dashed lines represent allowed crossings of bands with different symmetry eigenvalues. Only the black circles remain protected with spin-orbit coupling. At the (001) surface the symmetry is reduced to the symmorphic group $P4mm$. (b) Surface density of states for the effective model in Eqs.\,\ref{models} and \ref{modelns}, superposed with the bulk band structure at $k_z=0$ (dashed white line). (c) Surface density of states for a 26-band tight-binding model fitted to the DFT band structure of $\rm ZrSiS$. The Kramers' degeneracy at $\mathrm{\overline{X}}$ is lifted at the surface since the symmetry group is reduced to a symmorphic one. This creates a two-dimensional surface band. The effect is mostly visible in bands with large $k_z$ dispersion, in contrast to planar orbitals, such as $d_{x^2-y^2}$.}
    \label{fig:effective}
\end{figure*}
We develop an effective model Hamiltonian that qualitatively reproduces the observed surface density of states. 
We complement the symmetry discussion and the effective four band model Hamiltonian with a 26-band tight-binding model fitted to the DFT band structure of $\rm ZrSiS$. With this model, shown in Fig.\,\ref{fig:effective}, we can reproduce faithfully the surface density of states, as observed with ARPES in Fig.\,\ref{fig:SS_Overview} and effectively create a link between the model Hamiltonian and the {\em ab initio} DFT calculations.

Nonsymmorphic symmetries result from spatial transformations where no point is left invariant. They correspond to nontrivial extensions of point group transformations 
by fractional translations. 
These are screw axes and glide planes. 
A nonsymmorphic symmetry matrix representation $\bar g_\bk$ is momentum dependent, reflecting 
that when applied repeatedly it leads to a full lattice translation, not the identity. To avoid cluttering, we omit the momentum index
$\bk$. Nonsymmorphic symmetries are known to create `sticky points' in the electronic band structure \cite{Michel1999,Bradley1972}, 
 which are understood as the pinning of complex conjugate symmetry eigenvalues at time-reversal invariant momenta, due to time-reversal symmetry $\T$.
A common example of a nonsymmorphic space group is the tetragonal space group No.\ 129, $P4/nmm$, of which $\text{ZrSiS}$ is the a prominent example. It is characterized by a glide plane $\{M_z|\half\half0\}$, which is a mirror transformation in $\hat z$ and a half-translation along $\hat x+\hat y$; two screw axis $\{C_{2x}|\half00\}$ and $\{C_{2y}|0\half0\}$; a mirror $M_{xy}$ and inversion $\mathcal{I}$ symmetries. The combination of these symmetries enforces all bands to be degenerate at the X and M points (R and A points) in the bulk BZ, even in the presence of SOC \cite{Young2015}. In the absence of SOC, the enforced degeneracy is extended to the entire XM line. A schematic of the symmetry protected band degeneracies is shown in Fig.\,\ref{fig:effective}(a). Due to the generally small SOC in ZrSiS, we consider spinless electrons in our model. 
 We distinguish the enforced degeneracies (solid color), 
pinned to specific symmetry invariant momenta,  
from the crossing of bands with different symmetry eigenvalues along high symmetry planes
(dashed lines). The latter are responsible for the observed Dirac rings at $k_z=0,\pi$, and tubes at $k_x=0$ and $k_y=0$ \cite{Schoop2016}. The former is responsible for the four-fold nonsymmorphic Dirac points at X.
We can see representations of both cases in the bulk band structure of $\rm ZrSiS$ as depicted in Fig.\,\ref{fig:effective}(c). The states along the XM line are all four-fold degenerate due to the combination of a screw axis and time-reversal symmetries $\bar C_{2x}\T$. On the other hand, bands that transform under different eigenvalues of $\bar C_{2x}\mathcal{I}$ can cross 
along the $\Gamma$X line. The $k_z=0$ and $k_z=\pi$ planes, invariant under $\bar M_z$, allow for nodal rings as observed in Ref.\,\onlinecite{Schoop2016} (partially visible in Fig.\,\ref{fig:ARPES_FS}). The nodal ring is fixed to the Fermi level, as expected from the charge balance in $\rm ZrSiS$.

So far, the effect of the surface has not yet been taken into account. The crystal cleaves along the (001) plane breaking the translation symmetry along $\hat z$. As shown in the previous section, the surface introduces an asymmetry between the two nonsymmorphic sublattices ($\rm Zr$ atoms in the unit cell) due to their displacement along the $\hat z$ axis.  
As a consequence, \emph{all} nonsymmorphic symmetries are broken at the surface. The symmetry group is reduced to the spare wallpaper group $P4mm$ (space group No. 99), which only protects the crossing of surface bands along $\mathrm{\overline{\Gamma}\,\overline{X}}$ and $\mathrm{\overline{\Gamma}\,\overline{M}}$ (see Fig.\,\ref{fig:effective}(a), surface BZ). 
The remaining symmetry protection of surface bands along the $\mathrm{\overline{\Gamma}\,\overline{X}}$ line can be seen  directly in the ARPES data and slab DFT calculations in Fig.\,\ref{fig:SS_Overview}(a) and (b). In fact, from the bulk Dirac ring, only a surface Dirac point remains as $\bar M_z$ is broken. 
Most remarkable is the lifted degeneracy along $\mathrm{\overline{X}\,\overline{M}}$, which is protected in the bulk by the nonsymmorphic symmetries. The additional surface mass terms allow for a drastic gap opening which can detach a surface band from its parent bulk bands, creating a surface floating band.

To qualitatively simulate the band structure we consider a minimal spinless model without spin-orbit coupling, and with two sublattices A and B (eigenvalues of the Pauli matrix $\tau_z$), related by a fractional translation along $\hat x+\hat y$;  
and two orbitals, (eigenvalues of $\sigma_z$), even and odd with respect to $M_z$. 
The band Hamiltonian 
$
    \H=\sum_\bk\Phi^\dag_\bk H_\bk\Phi_\bk
$
acts on the basis $ \Phi_\bk=\ket{c_{A,+},c_{A,-},c_{B,+},c_{B,-}}_\bk$. Here, $c_{a,i}$ creates an electron in the sublattice $a=A,B$, with orbital $i=+,-$, even or odd under $\bar M_z$. 
 Imposing time-reversal and the spatial symmetries 
 we write the simple hopping Hamiltonian $H_\bk=H^{\rm s}_\bk+H^{\rm ns}_\bk$, with a symmorphic component 
 which preserves the sublattice,
\begin{align}
    H^{\rm s}_\bk=\mu+m\sigma_z+t^\pm_{xy}(\cos k_x+\cos k_y)\sigma_\pm+t^\pm_{z}\cos k_z\sigma_\pm,\label{models}
\end{align}
for $\sigma_\pm=1\pm\sigma_z$ the projector into the even(odd) orbital sectors; and a nonsymmorphic part 
\begin{align}
    H^{\rm ns}_\bk&=t(1+\cos k_x+\cos k_y+\cos (k_x+k_y))\tau_x\nonumber\\&+t(\sin k_x+\sin k_y+\sin (k_x+k_y))\tau_y,\label{modelns}
\end{align}
where the hopping coefficients, coupling the A and B sublattices, are fixed relative to each other \cite{zhao2016nonsymmorphic}. 

In Fig.\,\ref{fig:effective}(b) we show the bulk band structure of $H_\bk$ (in white dashed line), together with the surface density of states (color scale) in a slab calculation. We have  used $\mu=-0.1$, $m=0.5$, $t_{xy}^-=-t_{xy}^+=0.5$, $t_z^+=0.05$,  $t_z^-=0.02$ and $t=0.5$, to qualitatively correspond to the observed band structure in Fig.\,\ref{fig:SS_Overview}(a). 

The surface DOS is calculated following Ref.\,\onlinecite{Queiroz2015} taking a slab of 20 unit cells. We directly compare the effective model with the DFT-fitted 26 band tight-binding model in  Fig.\,\ref{fig:effective}(c), where we show the surface density of states. 
Motivated by the charge density distribution in Fig.\,\ref{fig:SS_Overview}(d), 
we expect the surface to create an unbalance between the local chemical potential and hopping amplitudes between the two sublattices.
This is expected from the modification in the crystal field at the surface.

\begin{figure}
    \centering
    \includegraphics[width=0.45\textwidth]{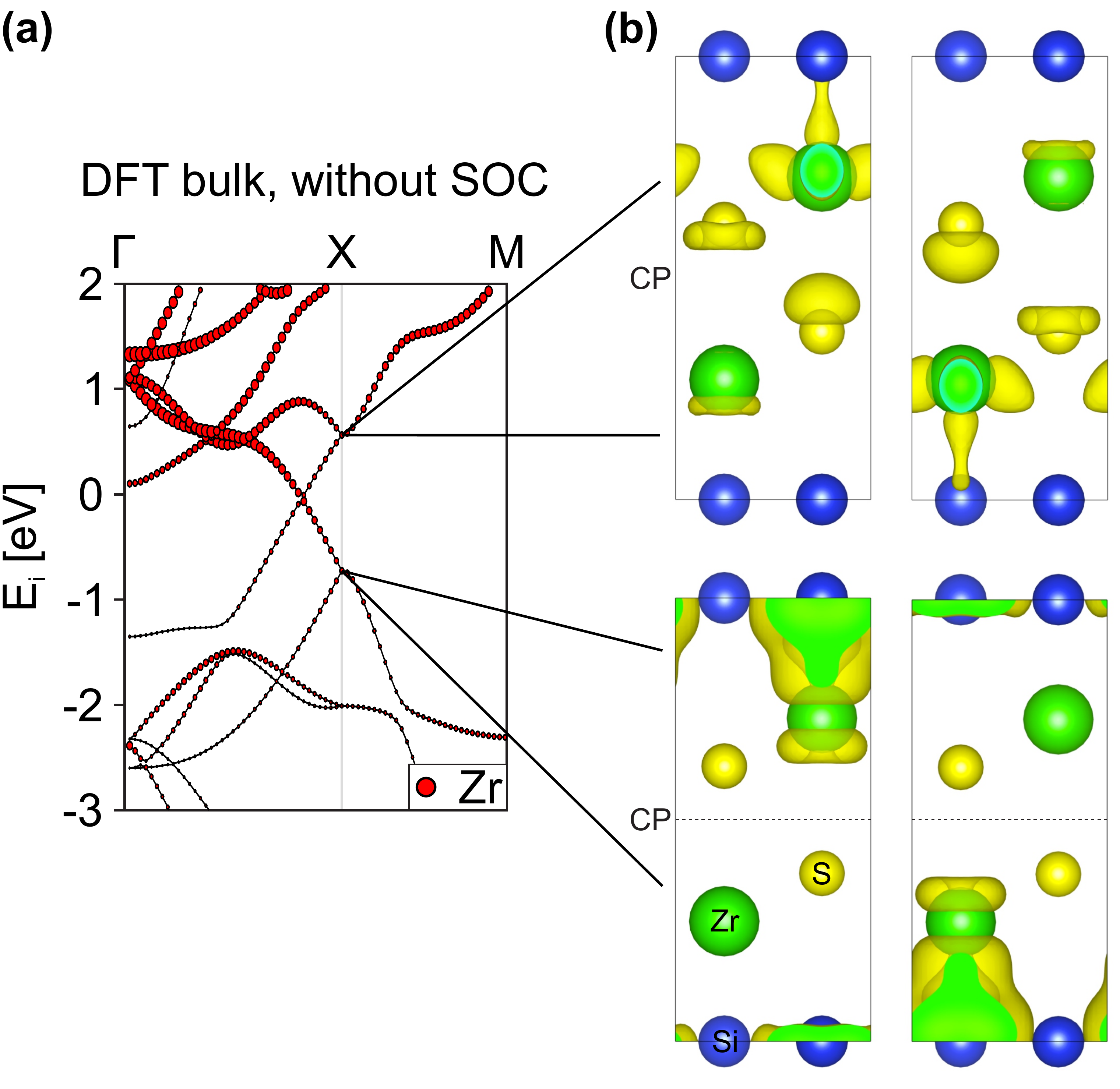}
    \caption{
    (a) DFT bulk calculation of $\rm ZrSiS$. The size of the red colored circles here stands for the overall Zr character in the unit cell, not just of the topmost one, as shown when considering the surface in Fig.\,\ref{fig:SS_Overview}(b) and (f). (b) Orbital character of the bulk bands at the X point shown through the isosurfaces of the charge density. The cleavage plane (CP) is shown as a dashed line in the unit cells. The color scale indicates the affiliation to the atoms (green - Zr, blue - Si, yellow - S). For the $\rm ZrSiS$ bulk material, the conduction band minimum at X shows a considerable DOS in the gap between the sulfur atoms, while the valence band maximum is much less affected by the introduction of a surface. 
    }
    \label{fig:DFT_X}
\end{figure}

We left the even orbital unaffected. This can be justified by the lack of charge density of the bulk band at the cleavage plane, and consequently less sensitivity (or even no sensitivity) to the symmetry reduction at the surface. This is the case in $\rm ZrSiS$, where the nonsymmorphic Dirac point below the Fermi level retains the bulk symmetry, since it does not occupy the Sulfur-$p_z$ orbitals at the surface layer. An explicit depiction of it can be found in Fig.\,\ref{fig:DFT_X}, where we show the real space charge density of the bulk states at the X point.
In the same figure, we can see that the Dirac point above the Fermi level has a significant Sulfur-$p_z$ contribution, which is highly affected by the presence of a surface, see Fig.\,\ref{fig:SS_Overview}(d). The symmetric orbital, deep in the material, is not affected. This breaks explicitly the symmetry that protects the degeneracy at X, leading to a floating, unpinned, surface band.


We simulate the symmetry reduction, by including a surface potential given by $H^{\rm surf}=-0.1(\tau_0+\tau_x)\sigma_+$ at the top and bottom layers. 
This term breaks the nonsymmorphic symmetry which requires intraband hopping to be equal in the two sublattices. 

\section{Response to surface modification} \label{sec:add}
In the previous sections, we have seen that the surface state results from a symmetry reduction at the surface due to a pronounced orbital deformation in the surface layer. In this context, we investigate how the surface state responds to a local chemical potential modification. For this purpose, we perform additional ARPES measurements with a monolayer of potassium evaporated on the surface.

In Fig.\,\ref{fig:SS_Overview}(e) and (f), we show the ARPES spectrum and DFT slab calculation along $\mathrm{\overline{\Gamma}\,\overline{X}\,\overline{M}}$, for a sample with a surface covered with 1\,ML potassium. The bulk bands remain nearly unmodified in comparison to the bare surface from Fig.\,\ref{fig:SS_Overview}(a) and (b); only a slight doping effect shifts the energy of the whole band structure. The states with surface character, in contrast, show a clear change. With potassium, the surface state shows a different band connectivity, one part following the bulk bands below the Fermi level along $\mathrm{\overline{X}\,\overline{M}}$ (SS1), while its other part going above the Fermi level seems to be connected to a second surface state (SS2). This change in connectivity is additional evidence for the trivial origin of the surface state.
The observed changes under potassium coating can be reproduced by performing DFT slab calculations of a surface covered with a regular array of potassium atoms. The results for a relaxed K distance of 2.74\,\AA~ from the surface are shown in Fig.\,\ref{fig:SS_Overview}(f). Again, we see very good agreement between theory and experiment. The two surface states, SS1 and SS2, which where connected previously (SS in Fig.\,\ref{fig:SS_Overview}(a) and (b)), are now gapped and connected to a band that roughly follows the bulk bands. In Fig.\,\ref{fig:SS_Overview}(b) this band already showed a similar orbital character as the surface state above, but not at the Fermi level, illustrated by the red circles. Without potassium, these states do not interfere and the surface state is crossing all bulk bands just above the nonsymmorphically protected point. With potassium, however, this band shows surface state character at the Fermi level leading to a gapping of the former crossing.

In that context it makes sense to analyze the orbital character in more detail to understand the change in the floating 2D band with potassium.
As explained in the previous theory section, the bands have different orbital character, allowing them to cross along the $\mathrm{\overline{\Gamma}\,\overline{X}}$ line. The potassium mixes their orbital character, therefore we observe an avoided crossing and a different band connectivity. This avoided crossing is also expected in the presence of spin-orbit coupling, as presented in Ref.\,\onlinecite{Schoop2016}. However, since SOC is very weak in $\rm ZrSiS$, the gap is only visible when the surface potential is varied by the potassium.

This prediction from the tight-binding model is very well caught in the DFT calculations in Fig.\,\ref{fig:SS_Overview}(f). However, ARPES does not capture the states that are presented in green in the DFT calculations. This is because they arise from a hypothetical regular array of potassium, which is not present in reality when evaporated at temperatures of 40\,K. 
Furthermore, we note that structural relaxation effects for the adsorbed K layer are another possible source for the discrepancy between DFT and ARPES experiments. Fig.\,\ref{fig:SS_Overview}(f) shows calculations for a relaxed K atom distance. The gap at the $\overline{X}$ point, between the surface states SS1 and SS2, is very pronounced and the change in the band connectivity is obvious. Nevertheless, this does not represent the measurements, where the gap is very small and the lower surface band even lies in the bulk bands. When increasing the distance of potassium atoms from the surface manually, the lower surface band shifts up, since the mixing in orbital character is reduced again. This implies a strong dependence of the DFT results on the modified surface potential agreeing with the fact that in the limit of a very large distance, a crossing is allowed again and the original band connectivity is restored. The overestimation of the gap in the DFT calculations can have different reasons. Firstly, a coverage of 1\,ML is experimentally not possible. In reality, an error of up to 10\,\% is not unlikely. A partial monolayer or a partial bilayer coverage of potassium might very well show a different relaxed distance from the surface. Secondly, the DFT calculation itself might overestimate the effect of the regular potassium layer on the surface. More accurate, beyond DFT approaches for the calculation of the band structure could provide additional insight and even better agreement with experiment. However, such methods are, due to their relatively large computational cost and the required system size for the surface model, currently not applicable.

\section{Conclusions} \label{sec:discussion}

We find that the origin of the surface states in $\rm ZrSiS$ and related compounds can be attributed to the symmetry reduction at the surface. This could be modeled by locally breaking the glide plane symmetry, introducing a large mass for the $\rm S$-$p_z$ and $\rm Zr$-$(d_{xz}+d_{yz})$ orbitals, which leads to a floating two-dimensional band close to the Fermi level. 
Therefore, it does originate neither from surface reconstruction nor from bulk topology. 
An explanation of the origin for the surface states on the basis of non-trivial topology or free standing monolayer as proposed in Ref.\,\onlinecite{Neupane2016ObservationZrSiS,lou2016emergence} can be excluded.

Due to the general symmetry arguments in our model, we conclude that 
similar surface floating bands 
are expected 
to appear in other layered nonsymmorphic compounds. Since a symmetry reduction at the surface lifts the degeneracy in the bulk band structure, materials with such `sticky points' are prone to show these states. Especially, other compounds from space group 129 are very likely to show them and indeed, e.g.~in $\rm ZrSiTe$ or $\rm HfSiS$ such unexpected surface states could have already been found \cite{topp2016non,Takane2016a}.

This does not necessarily imply that all degeneracies at high symmetry points are automatically lifted at the surface. In fact, the ARPES data in Fig.\,\ref{fig:SS_Overview}(a), as well as the tight-binding model in Fig.\,\ref{fig:effective}(c), show that only the upper degenerate point at $\mathrm{\overline{X}}$ is lifted, due to its orbital character and sufficient density of states in the cleavage plane.
This behavior has to be taken into account before applying this phenomenon to every nonsymmorphic degeneracy in the band structure of $\rm ZrSiS$ and related materials.

Lastly, we showed the peculiar properties of the surface states in question. They prevail for high photon energies up to 700\,eV and are prone to gapping around the $\mathrm{\overline{X}}$ point when covered with a monolayer of potassium, mixing the orbital characters of the states in question.


\section{Acknowledgements}
The authors thank Benjamin Balke for inspiring discussions and his insight into the matter of surface states. LM would like to thank Yang Zhang for providing the tight-binding parameters. Work at Argonne National Laboratory is supported by the U.S. Department of Energy, Office of Science, under Contract no.\,DE-AC02-06CH11357; additional support by National Science Foundation under Grant no.\,DMR-0703406. This work was partially supported by the DFG, 
proposal no.\,SCHO 1730/1-1.

\bibliographystyle{apsrev4-1}


\begin{thebibliography}{44}%
\makeatletter
\providecommand \@ifxundefined [1]{%
 \@ifx{#1\undefined}
}%
\providecommand \@ifnum [1]{%
 \ifnum #1\expandafter \@firstoftwo
 \else \expandafter \@secondoftwo
 \fi
}%
\providecommand \@ifx [1]{%
 \ifx #1\expandafter \@firstoftwo
 \else \expandafter \@secondoftwo
 \fi
}%
\providecommand \natexlab [1]{#1}%
\providecommand \enquote  [1]{``#1''}%
\providecommand \bibnamefont  [1]{#1}%
\providecommand \bibfnamefont [1]{#1}%
\providecommand \citenamefont [1]{#1}%
\providecommand \href@noop [0]{\@secondoftwo}%
\providecommand \href [0]{\begingroup \@sanitize@url \@href}%
\providecommand \@href[1]{\@@startlink{#1}\@@href}%
\providecommand \@@href[1]{\endgroup#1\@@endlink}%
\providecommand \@sanitize@url [0]{\catcode `\\12\catcode `\$12\catcode
  `\&12\catcode `\#12\catcode `\^12\catcode `\_12\catcode `\%12\relax}%
\providecommand \@@startlink[1]{}%
\providecommand \@@endlink[0]{}%
\providecommand \url  [0]{\begingroup\@sanitize@url \@url }%
\providecommand \@url [1]{\endgroup\@href {#1}{\urlprefix }}%
\providecommand \urlprefix  [0]{URL }%
\providecommand \Eprint [0]{\href }%
\providecommand \doibase [0]{http://dx.doi.org/}%
\providecommand \selectlanguage [0]{\@gobble}%
\providecommand \bibinfo  [0]{\@secondoftwo}%
\providecommand \bibfield  [0]{\@secondoftwo}%
\providecommand \translation [1]{[#1]}%
\providecommand \BibitemOpen [0]{}%
\providecommand \bibitemStop [0]{}%
\providecommand \bibitemNoStop [0]{.\EOS\space}%
\providecommand \EOS [0]{\spacefactor3000\relax}%
\providecommand \BibitemShut  [1]{\csname bibitem#1\endcsname}%
\let\auto@bib@innerbib\@empty
\bibitem [{\citenamefont {Kane}\ and\ \citenamefont
  {Mele}(2005)}]{Kane2005quantum}%
  \BibitemOpen
  \bibfield  {author} {\bibinfo {author} {\bibfnamefont {C.~L.}\ \bibnamefont
  {Kane}}\ and\ \bibinfo {author} {\bibfnamefont {E.~J.}\ \bibnamefont
  {Mele}},\ }\href {\doibase 10.1103/PhysRevLett.95.226801} {\bibfield
  {journal} {\bibinfo  {journal} {Physical Review Letters}\ }\textbf {\bibinfo
  {volume} {95}},\ \bibinfo {pages} {1} (\bibinfo {year} {2005})}\BibitemShut
  {NoStop}%
\bibitem [{\citenamefont {Fu}\ \emph {et~al.}(2007)\citenamefont {Fu},
  \citenamefont {Kane},\ and\ \citenamefont {Mele}}]{fu2007topological}%
  \BibitemOpen
  \bibfield  {author} {\bibinfo {author} {\bibfnamefont {L.}~\bibnamefont
  {Fu}}, \bibinfo {author} {\bibfnamefont {C.~L.}\ \bibnamefont {Kane}}, \ and\
  \bibinfo {author} {\bibfnamefont {E.~J.}\ \bibnamefont {Mele}},\ }\href@noop
  {} {\bibfield  {journal} {\bibinfo  {journal} {Physical Review Letters}\
  }\textbf {\bibinfo {volume} {98}},\ \bibinfo {pages} {106803} (\bibinfo
  {year} {2007})}\BibitemShut {NoStop}%
\bibitem [{\citenamefont {Schnyder}\ \emph {et~al.}(2008)\citenamefont
  {Schnyder}, \citenamefont {Ryu}, \citenamefont {Furusaki},\ and\
  \citenamefont {Ludwig}}]{schnyderPRB08}%
  \BibitemOpen
  \bibfield  {author} {\bibinfo {author} {\bibfnamefont {A.~P.}\ \bibnamefont
  {Schnyder}}, \bibinfo {author} {\bibfnamefont {S.}~\bibnamefont {Ryu}},
  \bibinfo {author} {\bibfnamefont {A.}~\bibnamefont {Furusaki}}, \ and\
  \bibinfo {author} {\bibfnamefont {A.~W.~W.}\ \bibnamefont {Ludwig}},\ }\href
  {\doibase 10.1103/PhysRevB.78.195125} {\bibfield  {journal} {\bibinfo
  {journal} {Physical Review B}\ }\textbf {\bibinfo {volume} {78}},\ \bibinfo
  {pages} {195125} (\bibinfo {year} {2008})}\BibitemShut {NoStop}%
\bibitem [{\citenamefont {Zhang}\ \emph {et~al.}(2009)\citenamefont {Zhang},
  \citenamefont {Liu}, \citenamefont {Qi}, \citenamefont {Dai}, \citenamefont
  {Fang},\ and\ \citenamefont {Zhang}}]{zhang2009topological}%
  \BibitemOpen
  \bibfield  {author} {\bibinfo {author} {\bibfnamefont {H.}~\bibnamefont
  {Zhang}}, \bibinfo {author} {\bibfnamefont {C.-X.}\ \bibnamefont {Liu}},
  \bibinfo {author} {\bibfnamefont {X.-L.}\ \bibnamefont {Qi}}, \bibinfo
  {author} {\bibfnamefont {X.}~\bibnamefont {Dai}}, \bibinfo {author}
  {\bibfnamefont {Z.}~\bibnamefont {Fang}}, \ and\ \bibinfo {author}
  {\bibfnamefont {S.-C.}\ \bibnamefont {Zhang}},\ }\href@noop {} {\bibfield
  {journal} {\bibinfo  {journal} {Nature physics}\ }\textbf {\bibinfo {volume}
  {5}},\ \bibinfo {pages} {438} (\bibinfo {year} {2009})}\BibitemShut {NoStop}%
\bibitem [{\citenamefont {Ando}(2013)}]{ando2013topological}%
  \BibitemOpen
  \bibfield  {author} {\bibinfo {author} {\bibfnamefont {Y.}~\bibnamefont
  {Ando}},\ }\href@noop {} {\bibfield  {journal} {\bibinfo  {journal} {Journal
  of the Physical Society of Japan}\ }\textbf {\bibinfo {volume} {82}},\
  \bibinfo {pages} {102001} (\bibinfo {year} {2013})}\BibitemShut {NoStop}%
\bibitem [{Nob(2016)}]{Nobel2016}%
  \BibitemOpen
  \href {https://www.nobelprize.org/nobel_prizes/physics/laureates/2016/}
  {\enquote {\bibinfo {title} {The nobel prize in physics 2016},}\ } (\bibinfo
  {year} {2016}),\ \bibinfo {note} {accessed: 2017-04-25,
  \url{https://www.nobelprize.org/nobel_prizes/physics/laureates/2016/}}\BibitemShut
  {NoStop}%
\bibitem [{\citenamefont {Wan}\ \emph {et~al.}(2011)\citenamefont {Wan},
  \citenamefont {Turner}, \citenamefont {Vishwanath},\ and\ \citenamefont
  {Savrasov}}]{wan2011topological}%
  \BibitemOpen
  \bibfield  {author} {\bibinfo {author} {\bibfnamefont {X.}~\bibnamefont
  {Wan}}, \bibinfo {author} {\bibfnamefont {A.~M.}\ \bibnamefont {Turner}},
  \bibinfo {author} {\bibfnamefont {A.}~\bibnamefont {Vishwanath}}, \ and\
  \bibinfo {author} {\bibfnamefont {S.~Y.}\ \bibnamefont {Savrasov}},\
  }\href@noop {} {\bibfield  {journal} {\bibinfo  {journal} {Physical Review
  B}\ }\textbf {\bibinfo {volume} {83}},\ \bibinfo {pages} {205101} (\bibinfo
  {year} {2011})}\BibitemShut {NoStop}%
\bibitem [{\citenamefont {Yan}\ and\ \citenamefont
  {Zhang}(2012)}]{yan2012topological}%
  \BibitemOpen
  \bibfield  {author} {\bibinfo {author} {\bibfnamefont {B.}~\bibnamefont
  {Yan}}\ and\ \bibinfo {author} {\bibfnamefont {S.-C.}\ \bibnamefont
  {Zhang}},\ }\href@noop {} {\bibfield  {journal} {\bibinfo  {journal} {Reports
  on Progress in Physics}\ }\textbf {\bibinfo {volume} {75}},\ \bibinfo {pages}
  {096501} (\bibinfo {year} {2012})}\BibitemShut {NoStop}%
\bibitem [{\citenamefont {Fang}\ \emph {et~al.}(2012)\citenamefont {Fang},
  \citenamefont {Gilbert}, \citenamefont {Dai},\ and\ \citenamefont
  {Bernevig}}]{Fang2012}%
  \BibitemOpen
  \bibfield  {author} {\bibinfo {author} {\bibfnamefont {C.}~\bibnamefont
  {Fang}}, \bibinfo {author} {\bibfnamefont {M.~J.}\ \bibnamefont {Gilbert}},
  \bibinfo {author} {\bibfnamefont {X.}~\bibnamefont {Dai}}, \ and\ \bibinfo
  {author} {\bibfnamefont {B.~A.}\ \bibnamefont {Bernevig}},\ }\href {\doibase
  10.1103/PhysRevLett.108.266802} {\bibfield  {journal} {\bibinfo  {journal}
  {Physical Review Letters}\ }\textbf {\bibinfo {volume} {108}},\ \bibinfo
  {pages} {1} (\bibinfo {year} {2012})}\BibitemShut {NoStop}%
\bibitem [{\citenamefont {Chiu}\ and\ \citenamefont
  {Schnyder}(2014)}]{Chiu2014a}%
  \BibitemOpen
  \bibfield  {author} {\bibinfo {author} {\bibfnamefont {C.-K.}\ \bibnamefont
  {Chiu}}\ and\ \bibinfo {author} {\bibfnamefont {A.~P.}\ \bibnamefont
  {Schnyder}},\ }\href {\doibase 10.1103/PhysRevB.90.205136} {\bibfield
  {journal} {\bibinfo  {journal} {Physical Review B}\ }\textbf {\bibinfo
  {volume} {90}},\ \bibinfo {pages} {205136} (\bibinfo {year}
  {2014})}\BibitemShut {NoStop}%
\bibitem [{\citenamefont {Liu}\ \emph {et~al.}(2014{\natexlab{a}})\citenamefont
  {Liu}, \citenamefont {Zhou}, \citenamefont {Zhang}, \citenamefont {Wang},
  \citenamefont {Weng}, \citenamefont {Prabhakaran}, \citenamefont {Mo},
  \citenamefont {Shen}, \citenamefont {Fang}, \citenamefont {Dai},
  \citenamefont {Hussain},\ and\ \citenamefont
  {Chen}}]{Liu2014DiscoveryNa3Bi.}%
  \BibitemOpen
  \bibfield  {author} {\bibinfo {author} {\bibfnamefont {Z.~K.}\ \bibnamefont
  {Liu}}, \bibinfo {author} {\bibfnamefont {B.}~\bibnamefont {Zhou}}, \bibinfo
  {author} {\bibfnamefont {Y.}~\bibnamefont {Zhang}}, \bibinfo {author}
  {\bibfnamefont {Z.~J.}\ \bibnamefont {Wang}}, \bibinfo {author}
  {\bibfnamefont {H.~M.}\ \bibnamefont {Weng}}, \bibinfo {author}
  {\bibfnamefont {D.}~\bibnamefont {Prabhakaran}}, \bibinfo {author}
  {\bibfnamefont {S.-K.}\ \bibnamefont {Mo}}, \bibinfo {author} {\bibfnamefont
  {Z.~X.}\ \bibnamefont {Shen}}, \bibinfo {author} {\bibfnamefont
  {Z.}~\bibnamefont {Fang}}, \bibinfo {author} {\bibfnamefont {X.}~\bibnamefont
  {Dai}}, \bibinfo {author} {\bibfnamefont {Z.}~\bibnamefont {Hussain}}, \ and\
  \bibinfo {author} {\bibfnamefont {Y.~L.}\ \bibnamefont {Chen}},\ }\href
  {\doibase 10.1126/science.1245085} {\bibfield  {journal} {\bibinfo  {journal}
  {Science (New York, N.Y.)}\ }\textbf {\bibinfo {volume} {343}} (\bibinfo
  {year} {2014}{\natexlab{a}}),\ 10.1126/science.1245085}\BibitemShut {NoStop}%
\bibitem [{\citenamefont {Liu}\ \emph {et~al.}(2014{\natexlab{b}})\citenamefont
  {Liu}, \citenamefont {Jiang}, \citenamefont {Zhou}, \citenamefont {Wang},
  \citenamefont {Zhang}, \citenamefont {Weng}, \citenamefont {Prabhakaran},
  \citenamefont {Mo}, \citenamefont {Peng}, \citenamefont {Dudin},
  \citenamefont {Kim}, \citenamefont {Hoesch}, \citenamefont {Fang},
  \citenamefont {Dai}, \citenamefont {Shen}, \citenamefont {Feng},
  \citenamefont {Hussain},\ and\ \citenamefont {Chen}}]{Liu2014}%
  \BibitemOpen
  \bibfield  {author} {\bibinfo {author} {\bibfnamefont {Z.~K.}\ \bibnamefont
  {Liu}}, \bibinfo {author} {\bibfnamefont {J.}~\bibnamefont {Jiang}}, \bibinfo
  {author} {\bibfnamefont {B.~B.}\ \bibnamefont {Zhou}}, \bibinfo {author}
  {\bibfnamefont {Z.~J.}\ \bibnamefont {Wang}}, \bibinfo {author}
  {\bibfnamefont {Y.}~\bibnamefont {Zhang}}, \bibinfo {author} {\bibfnamefont
  {H.~M.}\ \bibnamefont {Weng}}, \bibinfo {author} {\bibfnamefont
  {D.}~\bibnamefont {Prabhakaran}}, \bibinfo {author} {\bibfnamefont {S.-K.}\
  \bibnamefont {Mo}}, \bibinfo {author} {\bibfnamefont {H.}~\bibnamefont
  {Peng}}, \bibinfo {author} {\bibfnamefont {P.}~\bibnamefont {Dudin}},
  \bibinfo {author} {\bibfnamefont {T.}~\bibnamefont {Kim}}, \bibinfo {author}
  {\bibfnamefont {M.}~\bibnamefont {Hoesch}}, \bibinfo {author} {\bibfnamefont
  {Z.}~\bibnamefont {Fang}}, \bibinfo {author} {\bibfnamefont {X.}~\bibnamefont
  {Dai}}, \bibinfo {author} {\bibfnamefont {Z.-X.~X.}\ \bibnamefont {Shen}},
  \bibinfo {author} {\bibfnamefont {D.~L.}\ \bibnamefont {Feng}}, \bibinfo
  {author} {\bibfnamefont {Z.}~\bibnamefont {Hussain}}, \ and\ \bibinfo
  {author} {\bibfnamefont {Y.}~\bibnamefont {Chen}},\ }\href {\doibase
  10.1038/nmat3990} {\bibfield  {journal} {\bibinfo  {journal} {Nature
  Materials}\ }\textbf {\bibinfo {volume} {13}},\ \bibinfo {pages} {677}
  (\bibinfo {year} {2014}{\natexlab{b}})}\BibitemShut {NoStop}%
\bibitem [{\citenamefont {Shekhar}\ \emph {et~al.}(2015)\citenamefont
  {Shekhar}, \citenamefont {Nayak}, \citenamefont {Sun}, \citenamefont
  {Schmidt}, \citenamefont {Nicklas}, \citenamefont {Leermakers}, \citenamefont
  {Zeitler}, \citenamefont {Skourski}, \citenamefont {Wosnitza}, \citenamefont
  {Liu}, \citenamefont {Chen}, \citenamefont {Schnelle}, \citenamefont
  {Borrmann}, \citenamefont {Grin}, \citenamefont {Felser},\ and\ \citenamefont
  {Yan}}]{Shekhar2015a}%
  \BibitemOpen
  \bibfield  {author} {\bibinfo {author} {\bibfnamefont {C.}~\bibnamefont
  {Shekhar}}, \bibinfo {author} {\bibfnamefont {A.~K.}\ \bibnamefont {Nayak}},
  \bibinfo {author} {\bibfnamefont {Y.}~\bibnamefont {Sun}}, \bibinfo {author}
  {\bibfnamefont {M.}~\bibnamefont {Schmidt}}, \bibinfo {author} {\bibfnamefont
  {M.}~\bibnamefont {Nicklas}}, \bibinfo {author} {\bibfnamefont
  {I.}~\bibnamefont {Leermakers}}, \bibinfo {author} {\bibfnamefont
  {U.}~\bibnamefont {Zeitler}}, \bibinfo {author} {\bibfnamefont
  {Y.}~\bibnamefont {Skourski}}, \bibinfo {author} {\bibfnamefont
  {J.}~\bibnamefont {Wosnitza}}, \bibinfo {author} {\bibfnamefont
  {Z.}~\bibnamefont {Liu}}, \bibinfo {author} {\bibfnamefont {Y.}~\bibnamefont
  {Chen}}, \bibinfo {author} {\bibfnamefont {W.}~\bibnamefont {Schnelle}},
  \bibinfo {author} {\bibfnamefont {H.}~\bibnamefont {Borrmann}}, \bibinfo
  {author} {\bibfnamefont {Y.}~\bibnamefont {Grin}}, \bibinfo {author}
  {\bibfnamefont {C.}~\bibnamefont {Felser}}, \ and\ \bibinfo {author}
  {\bibfnamefont {B.}~\bibnamefont {Yan}},\ }\href {\doibase 10.1038/nphys3372}
  {\bibfield  {journal} {\bibinfo  {journal} {Nature Physics}\ }\textbf
  {\bibinfo {volume} {11}},\ \bibinfo {pages} {645} (\bibinfo {year}
  {2015})}\BibitemShut {NoStop}%
\bibitem [{\citenamefont {Chan}\ \emph {et~al.}(2016)\citenamefont {Chan},
  \citenamefont {Chiu}, \citenamefont {Chou},\ and\ \citenamefont
  {Schnyder}}]{Chan2016}%
  \BibitemOpen
  \bibfield  {author} {\bibinfo {author} {\bibfnamefont {Y.-H.}\ \bibnamefont
  {Chan}}, \bibinfo {author} {\bibfnamefont {C.-K.}\ \bibnamefont {Chiu}},
  \bibinfo {author} {\bibfnamefont {M.~Y.}\ \bibnamefont {Chou}}, \ and\
  \bibinfo {author} {\bibfnamefont {A.~P.}\ \bibnamefont {Schnyder}},\ }\href
  {\doibase 10.1103/PhysRevB.93.205132} {\bibfield  {journal} {\bibinfo
  {journal} {Physical Review B}\ }\textbf {\bibinfo {volume} {93}},\ \bibinfo
  {pages} {205132} (\bibinfo {year} {2016})}\BibitemShut {NoStop}%
\bibitem [{\citenamefont {Northrup}(1986)}]{northrup1986origin}%
  \BibitemOpen
  \bibfield  {author} {\bibinfo {author} {\bibfnamefont {J.~E.}\ \bibnamefont
  {Northrup}},\ }\href@noop {} {\bibfield  {journal} {\bibinfo  {journal}
  {Physical Review Letters}\ }\textbf {\bibinfo {volume} {57}},\ \bibinfo
  {pages} {154} (\bibinfo {year} {1986})}\BibitemShut {NoStop}%
\bibitem [{\citenamefont {Paggel}\ \emph {et~al.}(1999)\citenamefont {Paggel},
  \citenamefont {Miller},\ and\ \citenamefont {Chiang}}]{paggel1999quantum}%
  \BibitemOpen
  \bibfield  {author} {\bibinfo {author} {\bibfnamefont {J.}~\bibnamefont
  {Paggel}}, \bibinfo {author} {\bibfnamefont {T.}~\bibnamefont {Miller}}, \
  and\ \bibinfo {author} {\bibfnamefont {T.-C.}\ \bibnamefont {Chiang}},\
  }\href@noop {} {\bibfield  {journal} {\bibinfo  {journal} {Science}\ }\textbf
  {\bibinfo {volume} {283}},\ \bibinfo {pages} {1709} (\bibinfo {year}
  {1999})}\BibitemShut {NoStop}%
\bibitem [{\citenamefont {Davison}\ and\ \citenamefont
  {Steslicka}(1992)}]{Davison1992}%
  \BibitemOpen
  \bibfield  {author} {\bibinfo {author} {\bibfnamefont {S.~G.}\ \bibnamefont
  {Davison}}\ and\ \bibinfo {author} {\bibfnamefont {M.}~\bibnamefont
  {Steslicka}},\ }\href@noop {} {\emph {\bibinfo {title} {Basic Theory of
  Surface States -}}},\ \bibinfo {edition} {revised.}\ ed.\ (\bibinfo
  {publisher} {Clarendon Press},\ \bibinfo {address} {Oxford},\ \bibinfo {year}
  {1992})\BibitemShut {NoStop}%
\bibitem [{\citenamefont {Yan}\ \emph {et~al.}(2015)\citenamefont {Yan},
  \citenamefont {Stadtm{\"u}ller}, \citenamefont {Haag}, \citenamefont
  {Jakobs}, \citenamefont {Seidel}, \citenamefont {Jungkenn}, \citenamefont
  {Mathias}, \citenamefont {Cinchetti}, \citenamefont {Aeschlimann},\ and\
  \citenamefont {Felser}}]{yan2015topological}%
  \BibitemOpen
  \bibfield  {author} {\bibinfo {author} {\bibfnamefont {B.}~\bibnamefont
  {Yan}}, \bibinfo {author} {\bibfnamefont {B.}~\bibnamefont
  {Stadtm{\"u}ller}}, \bibinfo {author} {\bibfnamefont {N.}~\bibnamefont
  {Haag}}, \bibinfo {author} {\bibfnamefont {S.}~\bibnamefont {Jakobs}},
  \bibinfo {author} {\bibfnamefont {J.}~\bibnamefont {Seidel}}, \bibinfo
  {author} {\bibfnamefont {D.}~\bibnamefont {Jungkenn}}, \bibinfo {author}
  {\bibfnamefont {S.}~\bibnamefont {Mathias}}, \bibinfo {author} {\bibfnamefont
  {M.}~\bibnamefont {Cinchetti}}, \bibinfo {author} {\bibfnamefont
  {M.}~\bibnamefont {Aeschlimann}}, \ and\ \bibinfo {author} {\bibfnamefont
  {C.}~\bibnamefont {Felser}},\ }\href@noop {} {\bibfield  {journal} {\bibinfo
  {journal} {Nature communications}\ }\textbf {\bibinfo {volume} {6}},\
  \bibinfo {pages} {10167} (\bibinfo {year} {2015})}\BibitemShut {NoStop}%
\bibitem [{\citenamefont {Lindgren}\ and\ \citenamefont
  {Walld{\'e}n}(1979)}]{lindgren1979energy}%
  \BibitemOpen
  \bibfield  {author} {\bibinfo {author} {\bibfnamefont {S.}~\bibnamefont
  {Lindgren}}\ and\ \bibinfo {author} {\bibfnamefont {L.}~\bibnamefont
  {Walld{\'e}n}},\ }\href@noop {} {\bibfield  {journal} {\bibinfo  {journal}
  {Surface Science}\ }\textbf {\bibinfo {volume} {89}},\ \bibinfo {pages} {319}
  (\bibinfo {year} {1979})}\BibitemShut {NoStop}%
\bibitem [{\citenamefont {Tang}\ \emph {et~al.}(1993)\citenamefont {Tang},
  \citenamefont {Su},\ and\ \citenamefont {Heskett}}]{tang1993unoccupied}%
  \BibitemOpen
  \bibfield  {author} {\bibinfo {author} {\bibfnamefont {D.}~\bibnamefont
  {Tang}}, \bibinfo {author} {\bibfnamefont {C.}~\bibnamefont {Su}}, \ and\
  \bibinfo {author} {\bibfnamefont {D.}~\bibnamefont {Heskett}},\ }\href@noop
  {} {\bibfield  {journal} {\bibinfo  {journal} {Surface science}\ }\textbf
  {\bibinfo {volume} {295}},\ \bibinfo {pages} {427} (\bibinfo {year}
  {1993})}\BibitemShut {NoStop}%
\bibitem [{\citenamefont {Kevan}\ and\ \citenamefont
  {Gaylord}(1986)}]{kevan1986anomalous}%
  \BibitemOpen
  \bibfield  {author} {\bibinfo {author} {\bibfnamefont {S.}~\bibnamefont
  {Kevan}}\ and\ \bibinfo {author} {\bibfnamefont {R.}~\bibnamefont
  {Gaylord}},\ }\href@noop {} {\bibfield  {journal} {\bibinfo  {journal}
  {Physical Review Letters}\ }\textbf {\bibinfo {volume} {57}},\ \bibinfo
  {pages} {2975} (\bibinfo {year} {1986})}\BibitemShut {NoStop}%
\bibitem [{\citenamefont {Sandl}\ and\ \citenamefont
  {Bertel}(1994)}]{sandl1994surface}%
  \BibitemOpen
  \bibfield  {author} {\bibinfo {author} {\bibfnamefont {P.}~\bibnamefont
  {Sandl}}\ and\ \bibinfo {author} {\bibfnamefont {E.}~\bibnamefont {Bertel}},\
  }\href@noop {} {\bibfield  {journal} {\bibinfo  {journal} {Surface science}\
  }\textbf {\bibinfo {volume} {302}},\ \bibinfo {pages} {L325} (\bibinfo {year}
  {1994})}\BibitemShut {NoStop}%
\bibitem [{\citenamefont {Benia}\ \emph {et~al.}(2011)\citenamefont {Benia},
  \citenamefont {Lin}, \citenamefont {Kern},\ and\ \citenamefont
  {Ast}}]{benia2011reactive}%
  \BibitemOpen
  \bibfield  {author} {\bibinfo {author} {\bibfnamefont {H.~M.}\ \bibnamefont
  {Benia}}, \bibinfo {author} {\bibfnamefont {C.}~\bibnamefont {Lin}}, \bibinfo
  {author} {\bibfnamefont {K.}~\bibnamefont {Kern}}, \ and\ \bibinfo {author}
  {\bibfnamefont {C.~R.}\ \bibnamefont {Ast}},\ }\href@noop {} {\bibfield
  {journal} {\bibinfo  {journal} {Physical Review Letters}\ }\textbf {\bibinfo
  {volume} {107}},\ \bibinfo {pages} {177602} (\bibinfo {year}
  {2011})}\BibitemShut {NoStop}%
\bibitem [{\citenamefont {Cartier}\ \emph {et~al.}(1993)\citenamefont
  {Cartier}, \citenamefont {Stathis},\ and\ \citenamefont
  {Buchanan}}]{cartier1993passivation}%
  \BibitemOpen
  \bibfield  {author} {\bibinfo {author} {\bibfnamefont {E.}~\bibnamefont
  {Cartier}}, \bibinfo {author} {\bibfnamefont {J.}~\bibnamefont {Stathis}}, \
  and\ \bibinfo {author} {\bibfnamefont {D.}~\bibnamefont {Buchanan}},\
  }\href@noop {} {\bibfield  {journal} {\bibinfo  {journal} {Applied physics
  letters}\ }\textbf {\bibinfo {volume} {63}},\ \bibinfo {pages} {1510}
  (\bibinfo {year} {1993})}\BibitemShut {NoStop}%
\bibitem [{\citenamefont {Ast}\ \emph {et~al.}(2007)\citenamefont {Ast},
  \citenamefont {Henk}, \citenamefont {Ernst}, \citenamefont {Moreschini},
  \citenamefont {Falub}, \citenamefont {Pacil\'{e}}, \citenamefont {Bruno},
  \citenamefont {Kern},\ and\ \citenamefont {Grioni}}]{ast_giant_2007}%
  \BibitemOpen
  \bibfield  {author} {\bibinfo {author} {\bibfnamefont {C.~R.}\ \bibnamefont
  {Ast}}, \bibinfo {author} {\bibfnamefont {J.}~\bibnamefont {Henk}}, \bibinfo
  {author} {\bibfnamefont {A.}~\bibnamefont {Ernst}}, \bibinfo {author}
  {\bibfnamefont {L.}~\bibnamefont {Moreschini}}, \bibinfo {author}
  {\bibfnamefont {M.~C.}\ \bibnamefont {Falub}}, \bibinfo {author}
  {\bibfnamefont {D.}~\bibnamefont {Pacil\'{e}}}, \bibinfo {author}
  {\bibfnamefont {P.}~\bibnamefont {Bruno}}, \bibinfo {author} {\bibfnamefont
  {K.}~\bibnamefont {Kern}}, \ and\ \bibinfo {author} {\bibfnamefont
  {M.}~\bibnamefont {Grioni}},\ }\href {\doibase 10.1103/PhysRevLett.98.186807}
  {\bibfield  {journal} {\bibinfo  {journal} {Physical Review Letters}\
  }\textbf {\bibinfo {volume} {98}},\ \bibinfo {pages} {186807} (\bibinfo
  {year} {2007})}\BibitemShut {NoStop}%
\bibitem [{\citenamefont {Schoop}\ \emph {et~al.}(2016)\citenamefont {Schoop},
  \citenamefont {Ali}, \citenamefont {Stra{\ss}er}, \citenamefont {Topp},
  \citenamefont {Varykhalov}, \citenamefont {Marchenko}, \citenamefont
  {Duppel}, \citenamefont {Parkin}, \citenamefont {Lotsch},\ and\ \citenamefont
  {Ast}}]{Schoop2016}%
  \BibitemOpen
  \bibfield  {author} {\bibinfo {author} {\bibfnamefont {L.~M.}\ \bibnamefont
  {Schoop}}, \bibinfo {author} {\bibfnamefont {M.~N.}\ \bibnamefont {Ali}},
  \bibinfo {author} {\bibfnamefont {C.}~\bibnamefont {Stra{\ss}er}}, \bibinfo
  {author} {\bibfnamefont {A.}~\bibnamefont {Topp}}, \bibinfo {author}
  {\bibfnamefont {A.}~\bibnamefont {Varykhalov}}, \bibinfo {author}
  {\bibfnamefont {D.}~\bibnamefont {Marchenko}}, \bibinfo {author}
  {\bibfnamefont {V.}~\bibnamefont {Duppel}}, \bibinfo {author} {\bibfnamefont
  {S.}~\bibnamefont {Parkin}}, \bibinfo {author} {\bibfnamefont {B.~V.}\
  \bibnamefont {Lotsch}}, \ and\ \bibinfo {author} {\bibfnamefont {C.~R.}\
  \bibnamefont {Ast}},\ }\href {\doibase 10.1038/ncomms11696} {\bibfield
  {journal} {\bibinfo  {journal} {Nature Communications}\ }\textbf {\bibinfo
  {volume} {7}},\ \bibinfo {pages} {11696} (\bibinfo {year}
  {2016})}\BibitemShut {NoStop}%
\bibitem [{\citenamefont {Lv}\ \emph {et~al.}(2016)\citenamefont {Lv},
  \citenamefont {Zhang}, \citenamefont {Li}, \citenamefont {Yao}, \citenamefont
  {Chen}, \citenamefont {Zhou}, \citenamefont {Zhang}, \citenamefont {Lu},\
  and\ \citenamefont {Chen}}]{lv2016extremely}%
  \BibitemOpen
  \bibfield  {author} {\bibinfo {author} {\bibfnamefont {Y.-Y.}\ \bibnamefont
  {Lv}}, \bibinfo {author} {\bibfnamefont {B.-B.}\ \bibnamefont {Zhang}},
  \bibinfo {author} {\bibfnamefont {X.}~\bibnamefont {Li}}, \bibinfo {author}
  {\bibfnamefont {S.-H.}\ \bibnamefont {Yao}}, \bibinfo {author} {\bibfnamefont
  {Y.}~\bibnamefont {Chen}}, \bibinfo {author} {\bibfnamefont {J.}~\bibnamefont
  {Zhou}}, \bibinfo {author} {\bibfnamefont {S.-T.}\ \bibnamefont {Zhang}},
  \bibinfo {author} {\bibfnamefont {M.-H.}\ \bibnamefont {Lu}}, \ and\ \bibinfo
  {author} {\bibfnamefont {Y.-F.}\ \bibnamefont {Chen}},\ }\href@noop {}
  {\bibfield  {journal} {\bibinfo  {journal} {Applied Physics Letters}\
  }\textbf {\bibinfo {volume} {108}},\ \bibinfo {pages} {244101} (\bibinfo
  {year} {2016})}\BibitemShut {NoStop}%
\bibitem [{\citenamefont {Ali}\ \emph {et~al.}(2016)\citenamefont {Ali},
  \citenamefont {Schoop}, \citenamefont {Garg}, \citenamefont {Lippmann},
  \citenamefont {Lara}, \citenamefont {Lotsch},\ and\ \citenamefont
  {Parkin}}]{ali2016butterfly}%
  \BibitemOpen
  \bibfield  {author} {\bibinfo {author} {\bibfnamefont {M.~N.}\ \bibnamefont
  {Ali}}, \bibinfo {author} {\bibfnamefont {L.~M.}\ \bibnamefont {Schoop}},
  \bibinfo {author} {\bibfnamefont {C.}~\bibnamefont {Garg}}, \bibinfo {author}
  {\bibfnamefont {J.~M.}\ \bibnamefont {Lippmann}}, \bibinfo {author}
  {\bibfnamefont {E.}~\bibnamefont {Lara}}, \bibinfo {author} {\bibfnamefont
  {B.}~\bibnamefont {Lotsch}}, \ and\ \bibinfo {author} {\bibfnamefont {S.~S.}\
  \bibnamefont {Parkin}},\ }\href@noop {} {\bibfield  {journal} {\bibinfo
  {journal} {Science advances}\ }\textbf {\bibinfo {volume} {2}},\ \bibinfo
  {pages} {e1601742} (\bibinfo {year} {2016})}\BibitemShut {NoStop}%
\bibitem [{\citenamefont {Topp}\ \emph {et~al.}(2016)\citenamefont {Topp},
  \citenamefont {Lippmann}, \citenamefont {Varykhalov}, \citenamefont {Duppel},
  \citenamefont {Lotsch}, \citenamefont {Ast},\ and\ \citenamefont
  {Schoop}}]{topp2016non}%
  \BibitemOpen
  \bibfield  {author} {\bibinfo {author} {\bibfnamefont {A.}~\bibnamefont
  {Topp}}, \bibinfo {author} {\bibfnamefont {J.~M.}\ \bibnamefont {Lippmann}},
  \bibinfo {author} {\bibfnamefont {A.}~\bibnamefont {Varykhalov}}, \bibinfo
  {author} {\bibfnamefont {V.}~\bibnamefont {Duppel}}, \bibinfo {author}
  {\bibfnamefont {B.~V.}\ \bibnamefont {Lotsch}}, \bibinfo {author}
  {\bibfnamefont {C.~R.}\ \bibnamefont {Ast}}, \ and\ \bibinfo {author}
  {\bibfnamefont {L.~M.}\ \bibnamefont {Schoop}},\ }\href@noop {} {\bibfield
  {journal} {\bibinfo  {journal} {New Journal of Physics}\ }\textbf {\bibinfo
  {volume} {18}},\ \bibinfo {pages} {125014} (\bibinfo {year}
  {2016})}\BibitemShut {NoStop}%
\bibitem [{\citenamefont {Takane}\ \emph {et~al.}(2016)\citenamefont {Takane},
  \citenamefont {Wang}, \citenamefont {Souma}, \citenamefont {Nakayama},
  \citenamefont {Trang}, \citenamefont {Sato}, \citenamefont {Takahashi},\ and\
  \citenamefont {Ando}}]{Takane2016a}%
  \BibitemOpen
  \bibfield  {author} {\bibinfo {author} {\bibfnamefont {D.}~\bibnamefont
  {Takane}}, \bibinfo {author} {\bibfnamefont {Z.}~\bibnamefont {Wang}},
  \bibinfo {author} {\bibfnamefont {S.}~\bibnamefont {Souma}}, \bibinfo
  {author} {\bibfnamefont {K.}~\bibnamefont {Nakayama}}, \bibinfo {author}
  {\bibfnamefont {C.~X.}\ \bibnamefont {Trang}}, \bibinfo {author}
  {\bibfnamefont {T.}~\bibnamefont {Sato}}, \bibinfo {author} {\bibfnamefont
  {T.}~\bibnamefont {Takahashi}}, \ and\ \bibinfo {author} {\bibfnamefont
  {Y.}~\bibnamefont {Ando}},\ }\href {\doibase 10.1103/PhysRevB.94.121108}
  {\bibfield  {journal} {\bibinfo  {journal} {Physical Review B}\ }\textbf
  {\bibinfo {volume} {94}},\ \bibinfo {pages} {121108} (\bibinfo {year}
  {2016})}\BibitemShut {NoStop}%
\bibitem [{\citenamefont {Xu}\ \emph {et~al.}(2015)\citenamefont {Xu},
  \citenamefont {Song}, \citenamefont {Nie}, \citenamefont {Weng},
  \citenamefont {Fang},\ and\ \citenamefont {Dai}}]{xu2015two}%
  \BibitemOpen
  \bibfield  {author} {\bibinfo {author} {\bibfnamefont {Q.}~\bibnamefont
  {Xu}}, \bibinfo {author} {\bibfnamefont {Z.}~\bibnamefont {Song}}, \bibinfo
  {author} {\bibfnamefont {S.}~\bibnamefont {Nie}}, \bibinfo {author}
  {\bibfnamefont {H.}~\bibnamefont {Weng}}, \bibinfo {author} {\bibfnamefont
  {Z.}~\bibnamefont {Fang}}, \ and\ \bibinfo {author} {\bibfnamefont
  {X.}~\bibnamefont {Dai}},\ }\href@noop {} {\bibfield  {journal} {\bibinfo
  {journal} {Physical Review B}\ }\textbf {\bibinfo {volume} {92}},\ \bibinfo
  {pages} {205310} (\bibinfo {year} {2015})}\BibitemShut {NoStop}%
\bibitem [{\citenamefont {Neupane}\ \emph {et~al.}(2016)\citenamefont
  {Neupane}, \citenamefont {Belopolski}, \citenamefont {Hosen}, \citenamefont
  {Sanchez}, \citenamefont {Sankar}, \citenamefont {Szlawska}, \citenamefont
  {Xu}, \citenamefont {Dimitri}, \citenamefont {Dhakal}, \citenamefont
  {Maldonado}, \citenamefont {Oppeneer}, \citenamefont {Kaczorowski},
  \citenamefont {Chou}, \citenamefont {Hasan},\ and\ \citenamefont
  {Durakiewicz}}]{Neupane2016ObservationZrSiS}%
  \BibitemOpen
  \bibfield  {author} {\bibinfo {author} {\bibfnamefont {M.}~\bibnamefont
  {Neupane}}, \bibinfo {author} {\bibfnamefont {I.}~\bibnamefont {Belopolski}},
  \bibinfo {author} {\bibfnamefont {M.~M.}\ \bibnamefont {Hosen}}, \bibinfo
  {author} {\bibfnamefont {D.~S.}\ \bibnamefont {Sanchez}}, \bibinfo {author}
  {\bibfnamefont {R.}~\bibnamefont {Sankar}}, \bibinfo {author} {\bibfnamefont
  {M.}~\bibnamefont {Szlawska}}, \bibinfo {author} {\bibfnamefont {S.-Y.}\
  \bibnamefont {Xu}}, \bibinfo {author} {\bibfnamefont {K.}~\bibnamefont
  {Dimitri}}, \bibinfo {author} {\bibfnamefont {N.}~\bibnamefont {Dhakal}},
  \bibinfo {author} {\bibfnamefont {P.}~\bibnamefont {Maldonado}}, \bibinfo
  {author} {\bibfnamefont {P.~M.}\ \bibnamefont {Oppeneer}}, \bibinfo {author}
  {\bibfnamefont {D.}~\bibnamefont {Kaczorowski}}, \bibinfo {author}
  {\bibfnamefont {F.}~\bibnamefont {Chou}}, \bibinfo {author} {\bibfnamefont
  {M.~Z.}\ \bibnamefont {Hasan}}, \ and\ \bibinfo {author} {\bibfnamefont
  {T.}~\bibnamefont {Durakiewicz}},\ }\href {\doibase
  10.1103/PhysRevB.93.201104} {\bibfield  {journal} {\bibinfo  {journal}
  {Physical Review B}\ }\textbf {\bibinfo {volume} {93}},\ \bibinfo {pages}
  {201104} (\bibinfo {year} {2016})}\BibitemShut {NoStop}%
\bibitem [{\citenamefont {Lou}\ \emph {et~al.}(2016)\citenamefont {Lou},
  \citenamefont {Ma}, \citenamefont {Xu}, \citenamefont {Fu}, \citenamefont
  {Kong}, \citenamefont {Shi}, \citenamefont {Richard}, \citenamefont {Weng},
  \citenamefont {Fang}, \citenamefont {Sun} \emph {et~al.}}]{lou2016emergence}%
  \BibitemOpen
  \bibfield  {author} {\bibinfo {author} {\bibfnamefont {R.}~\bibnamefont
  {Lou}}, \bibinfo {author} {\bibfnamefont {J.-Z.}\ \bibnamefont {Ma}},
  \bibinfo {author} {\bibfnamefont {Q.-N.}\ \bibnamefont {Xu}}, \bibinfo
  {author} {\bibfnamefont {B.-B.}\ \bibnamefont {Fu}}, \bibinfo {author}
  {\bibfnamefont {L.-Y.}\ \bibnamefont {Kong}}, \bibinfo {author}
  {\bibfnamefont {Y.-G.}\ \bibnamefont {Shi}}, \bibinfo {author} {\bibfnamefont
  {P.}~\bibnamefont {Richard}}, \bibinfo {author} {\bibfnamefont {H.-M.}\
  \bibnamefont {Weng}}, \bibinfo {author} {\bibfnamefont {Z.}~\bibnamefont
  {Fang}}, \bibinfo {author} {\bibfnamefont {S.-S.}\ \bibnamefont {Sun}},
  \emph {et~al.},\ }\href@noop {} {\bibfield  {journal} {\bibinfo  {journal}
  {Physical Review B}\ }\textbf {\bibinfo {volume} {93}},\ \bibinfo {pages}
  {241104} (\bibinfo {year} {2016})}\BibitemShut {NoStop}%
\bibitem [{\citenamefont {Wang}\ \emph {et~al.}(2016)\citenamefont {Wang},
  \citenamefont {Alexandradinata}, \citenamefont {Cava},\ and\ \citenamefont
  {Bernevig}}]{wang2016hourglass}%
  \BibitemOpen
  \bibfield  {author} {\bibinfo {author} {\bibfnamefont {Z.}~\bibnamefont
  {Wang}}, \bibinfo {author} {\bibfnamefont {A.}~\bibnamefont
  {Alexandradinata}}, \bibinfo {author} {\bibfnamefont {R.}~\bibnamefont
  {Cava}}, \ and\ \bibinfo {author} {\bibfnamefont {B.~A.}\ \bibnamefont
  {Bernevig}},\ }\href@noop {} {\bibfield  {journal} {\bibinfo  {journal}
  {Nature}\ }\textbf {\bibinfo {volume} {532}},\ \bibinfo {pages} {189}
  (\bibinfo {year} {2016})}\BibitemShut {NoStop}%
\bibitem [{\citenamefont {Wieder}\ \emph {et~al.}(2017)\citenamefont {Wieder},
  \citenamefont {Bradlyn}, \citenamefont {Wang}, \citenamefont {Cano},
  \citenamefont {Kim}, \citenamefont {Kim}, \citenamefont {Rappe},
  \citenamefont {Kane},\ and\ \citenamefont {Bernevig}}]{wieder2017wallpaper}%
  \BibitemOpen
  \bibfield  {author} {\bibinfo {author} {\bibfnamefont {B.~J.}\ \bibnamefont
  {Wieder}}, \bibinfo {author} {\bibfnamefont {B.}~\bibnamefont {Bradlyn}},
  \bibinfo {author} {\bibfnamefont {Z.}~\bibnamefont {Wang}}, \bibinfo {author}
  {\bibfnamefont {J.}~\bibnamefont {Cano}}, \bibinfo {author} {\bibfnamefont
  {Y.}~\bibnamefont {Kim}}, \bibinfo {author} {\bibfnamefont {H.-S.~D.}\
  \bibnamefont {Kim}}, \bibinfo {author} {\bibfnamefont {A.}~\bibnamefont
  {Rappe}}, \bibinfo {author} {\bibfnamefont {C.}~\bibnamefont {Kane}}, \ and\
  \bibinfo {author} {\bibfnamefont {B.~A.}\ \bibnamefont {Bernevig}},\
  }\href@noop {} {\bibfield  {journal} {\bibinfo  {journal} {arXiv preprint
  arXiv:1705.01617}\ } (\bibinfo {year} {2017})}\BibitemShut {NoStop}%
\bibitem [{\citenamefont {Kresse}\ and\ \citenamefont
  {Joubert}(1999)}]{kresse1999}%
  \BibitemOpen
  \bibfield  {author} {\bibinfo {author} {\bibfnamefont {G.}~\bibnamefont
  {Kresse}}\ and\ \bibinfo {author} {\bibfnamefont {D.}~\bibnamefont
  {Joubert}},\ }\href {\doibase 10.1103/PhysRevB.59.1758} {\bibfield  {journal}
  {\bibinfo  {journal} {Phys. Rev. B}\ }\textbf {\bibinfo {volume} {59}},\
  \bibinfo {pages} {1758} (\bibinfo {year} {1999})}\BibitemShut {NoStop}%
\bibitem [{Note1()}]{Note1}%
  \BibitemOpen
  \bibinfo {note} {See VASP manual: www.vasp.at .}\BibitemShut {Stop}%
\bibitem [{\citenamefont {Hsieh}\ \emph {et~al.}(1987)\citenamefont {Hsieh},
  \citenamefont {John}, \citenamefont {Miller},\ and\ \citenamefont
  {Chiang}}]{hsieh_resonances_1987}%
  \BibitemOpen
  \bibfield  {author} {\bibinfo {author} {\bibfnamefont {T.~C.}\ \bibnamefont
  {Hsieh}}, \bibinfo {author} {\bibfnamefont {P.}~\bibnamefont {John}},
  \bibinfo {author} {\bibfnamefont {T.}~\bibnamefont {Miller}}, \ and\ \bibinfo
  {author} {\bibfnamefont {T.-C.}\ \bibnamefont {Chiang}},\ }\href {\doibase
  10.1103/PhysRevB.35.3728} {\bibfield  {journal} {\bibinfo  {journal}
  {Physical Review B}\ }\textbf {\bibinfo {volume} {35}},\ \bibinfo {pages}
  {3728} (\bibinfo {year} {1987})}\BibitemShut {NoStop}%
\bibitem [{\citenamefont {Lobo-Checa}\ \emph {et~al.}(2011)\citenamefont
  {Lobo-Checa}, \citenamefont {Ortega}, \citenamefont {Mascaraque},
  \citenamefont {Michel},\ and\ \citenamefont
  {Krasovskii}}]{lobo-checa_effect_2011}%
  \BibitemOpen
  \bibfield  {author} {\bibinfo {author} {\bibfnamefont {J.}~\bibnamefont
  {Lobo-Checa}}, \bibinfo {author} {\bibfnamefont {J.~E.}\ \bibnamefont
  {Ortega}}, \bibinfo {author} {\bibfnamefont {A.}~\bibnamefont {Mascaraque}},
  \bibinfo {author} {\bibfnamefont {E.~G.}\ \bibnamefont {Michel}}, \ and\
  \bibinfo {author} {\bibfnamefont {E.~E.}\ \bibnamefont {Krasovskii}},\ }\href
  {\doibase 10.1103/PhysRevB.84.245419} {\bibfield  {journal} {\bibinfo
  {journal} {Physical Review B}\ }\textbf {\bibinfo {volume} {84}},\ \bibinfo
  {pages} {245419} (\bibinfo {year} {2011})}\BibitemShut {NoStop}%
\bibitem [{\citenamefont {Michel}\ and\ \citenamefont
  {Zak}(1999)}]{Michel1999}%
  \BibitemOpen
  \bibfield  {author} {\bibinfo {author} {\bibfnamefont {L.}~\bibnamefont
  {Michel}}\ and\ \bibinfo {author} {\bibfnamefont {J.}~\bibnamefont {Zak}},\
  }\href {\doibase 10.1103/PhysRevB.59.5998} {\bibfield  {journal} {\bibinfo
  {journal} {Physical Review B}\ }\textbf {\bibinfo {volume} {59}},\ \bibinfo
  {pages} {5998} (\bibinfo {year} {1999})}\BibitemShut {NoStop}%
\bibitem [{\citenamefont {Bradley}\ and\ \citenamefont
  {Cracknell}(1972)}]{Bradley1972}%
  \BibitemOpen
  \bibfield  {author} {\bibinfo {author} {\bibfnamefont {C.~J.}\ \bibnamefont
  {Bradley}}\ and\ \bibinfo {author} {\bibfnamefont {A.}~\bibnamefont
  {Cracknell}},\ }\href {\doibase 10.5840/bradley19984110} {\emph {\bibinfo
  {title} {{The Mathematical Theory of Symmetry in Solids}}}}\ (\bibinfo
  {publisher} {Oxford University Press},\ \bibinfo {year} {1972})\BibitemShut
  {NoStop}%
\bibitem [{\citenamefont {Young}\ and\ \citenamefont {Kane}(2015)}]{Young2015}%
  \BibitemOpen
  \bibfield  {author} {\bibinfo {author} {\bibfnamefont {S.~M.}\ \bibnamefont
  {Young}}\ and\ \bibinfo {author} {\bibfnamefont {C.~L.}\ \bibnamefont
  {Kane}},\ }\href {\doibase 10.1103/PhysRevLett.115.126803} {\bibfield
  {journal} {\bibinfo  {journal} {Physical Review Letters}\ }\textbf {\bibinfo
  {volume} {115}},\ \bibinfo {pages} {1} (\bibinfo {year} {2015})}\BibitemShut
  {NoStop}%
\bibitem [{\citenamefont {Zhao}\ and\ \citenamefont
  {Schnyder}(2016)}]{zhao2016nonsymmorphic}%
  \BibitemOpen
  \bibfield  {author} {\bibinfo {author} {\bibfnamefont {Y.}~\bibnamefont
  {Zhao}}\ and\ \bibinfo {author} {\bibfnamefont {A.~P.}\ \bibnamefont
  {Schnyder}},\ }\href@noop {} {\bibfield  {journal} {\bibinfo  {journal}
  {Physical Review B}\ }\textbf {\bibinfo {volume} {94}},\ \bibinfo {pages}
  {195109} (\bibinfo {year} {2016})}\BibitemShut {NoStop}%
\bibitem [{\citenamefont {Queiroz}\ and\ \citenamefont
  {Schnyder}(2015)}]{Queiroz2015}%
  \BibitemOpen
  \bibfield  {author} {\bibinfo {author} {\bibfnamefont {R.}~\bibnamefont
  {Queiroz}}\ and\ \bibinfo {author} {\bibfnamefont {A.~P.}\ \bibnamefont
  {Schnyder}},\ }\href {\doibase 10.1103/PhysRevB.91.014202} {\bibfield
  {journal} {\bibinfo  {journal} {Physical Review B}\ }\textbf {\bibinfo
  {volume} {91}},\ \bibinfo {pages} {014202} (\bibinfo {year}
  {2015})}\BibitemShut {NoStop}%
\end{thebibliography}

%

 \end{document}